\documentclass[a4paper,11pt,draftcls,journal,onecolumn]{IEEEtran}

\usepackage{graphicx,amsfonts,latexsym}
\usepackage{amssymb,amsmath}

\usepackage{algpseudocode,algorithm}

\usepackage{color}
\usepackage{upquote}
\usepackage[normalem]{ulem}

\newtheorem {example}{Example}
\newtheorem {remark}{Remark}
\newtheorem{theorem}{Theorem}

\newtheorem {assumption}{Assumption}

\makeatletter
\renewcommand*{\@opargbegintheorem}[3]{\trivlist
      \item[\hskip \labelsep{\itshape #1\ #2}] \textit{(#3)}\ }
\makeatother

\newcommand{\bit}{\begin{itemize}}
\newcommand{\eit}{\end{itemize}}
\newcommand{\ben}{\begin{enumerate}}
\newcommand{\een}{\end{enumerate}}

\newcommand{\beq}{\begin{equation}}
\newcommand{\eeq}{\end{equation}}
\newcommand{\beas}{\begin{eqnarray*}}
\newcommand{\eeas}{\end{eqnarray*}}
\newcommand{\bea}{\begin{eqnarray}}
\newcommand{\eea}{\end{eqnarray}}
\newcommand{\ba}{\begin{array}}
\newcommand{\ea}{\end{array}}
\newcommand{\qed}{\hfill $\square$}

\newcommand{\q}{{\eta}}
\newcommand{\Q}{\mathcal{N}}

\newcommand{\I}{\mathbb{I}}
\newcommand{\ind}[1]{\I\left(#1\right)}
\newcommand{\vol}[1]{\mathrm{vol}\left[#1\right]}
\newcommand{\Eval}[1]{\mathbb{E}\left[#1\right]}
\newcommand{\uni}{\mathbf{1}}
\newcommand{\tran}{^{\top}\!}
\newcommand{\erem}{\hfill $\diamond$}

\newcommand{\cA}{{\cal A}}
\newcommand{\cB}{{\cal B}}
\newcommand{\cC}{{\cal C}}

\newcommand{\cE}{{\cal E}}

\newcommand{\cH}{{\cal H}}
\newcommand{\cI}{{\cal I}}

\newcommand{\cP}{{\cal P}}

\newcommand{\cS}{{\cal S}}

\newcommand{\cU}{{\cal U}}

\newcommand{\cX}{{\cal X}}

\newcommand{\Real}[1]{ { {\mathbb R}^{#1} } }

\newcommand{\ped}[1]{_{\mathrm{#1}}}
\newcommand{\apex}[1]{^{\mathrm{#1}}}

\newcommand{\muQ}{{\mu}_{\Q}}

\newcommand{\muU}[1]{\lambda_{#1}}
\newcommand{\pU}[1]{{\cU_{#1}}}

\newcommand{\muX}{\tilde{\mu}_{\Ix}}
\newcommand{\muZ}{\tilde{\mu}_{\cS\!\Ix}}

\newcommand{\Xe}{ \mathcal{X}_\epsilon }

\newcommand{\Cyl}{ \cC}
\newcommand{\vstar}{ \phi\ped{o}}

\newcommand{\Ix}{{\cI^{-1}_{y}} }
\newcommand{\Prob}[1]{\mathrm{Prob}\left\{#1\right\}}

\newcommand{\Pmax}{\textbf{(P-max-int) }}

\newcommand{\proof}{\vspace{0.2cm} \noindent \textit{Proof: }}
\renewcommand{\qed}{{\hfill $\square$} \vspace{.4cm}}

\newcommand{\Vcyl}{\mathrm{V}_{\cC}}
\newcommand{\tS}{ \bar{\cS}}

\begin{document}
\title{\huge Probabilistic Optimal Estimation under Uncertainty}
\author{\textsc{Fabrizio Dabbene,\quad  Mario Sznaier,\quad and\quad Roberto Tempo$^*$}
\thanks{$^*$ Corresponding author}
\thanks{F. Dabbene and R. Tempo are with the CNR-IEIIT Institute, Politecnico di Torino, Italy (e-mail: fabrizio.dabbene@polito.it, roberto.tempo@polito.it).}
\thanks{M. Sznaier is with the Northeastern University, Boston, MA 02115, USA (e-mail: msznaier@ece.neu.edu).}
}
\maketitle

\begin{abstract}
The classical approach to system identification is based on stochastic assumptions about the measurement error, and provides estimates that have random nature. Worst-case identification, on the other hand, only assumes the knowledge of deterministic error bounds, and establishes guaranteed
estimates, thus being in principle better suited for the use in control
design. However, a main limitation
of such deterministic bounds lies on their potential conservatism, thus leading to estimates of restricted use.

In this paper, we propose a rapprochement between the stochastic and worst-case paradigms.
In particular, based on a probabilistic framework for linear estimation problems, we derive new computational results. These results combine elements from information-based complexity with
recent developments in the theory of randomized algorithms. 
The main idea in this line of research is to ``discard" sets of measure at most $\epsilon$, where $\epsilon$ is a probabilistic  accuracy, from the set of deterministic estimates. Therefore, we are decreasing the
so-called worst-case radius of information at the expense of a given probabilistic  ``risk." 

In this setting, we compute a trade-off curve, called \textit{violation function}, which  shows how the
radius of information decreases as a function of the accuracy.  To this end, we construct  randomized and deterministic algorithms which provide approximations of
this function. We report extensive simulations showing numerical comparisons
between the stochastic, worst-case and probabilistic approaches, thus demonstrating the efficacy of the methods proposed in this paper.

\end{abstract}


\clearpage

\section{Introduction and Preliminaries}
\label{sec-intro}


The mainstream paradigm for system identification is the classical stochastic approach, 
 see \cite{Ljung:99} and the special issues \cite{LjuVic:05,SVWW:05}, which has been very successful also in many applications, such as e.g.\ process control and systems biology.
This approach assumes that the available observations are contaminated by {\em random} noise normally distributed, and has the goal to derive soft bounds on the estimation errors.  In this setting, optimality is guaranteed  in a probabilistic sense and the resulting algorithms often enjoy convergence properties only asymptotically. 

In the last decades, several authors focused their attention on the so-called set-membership identification which aims at the computation of hard bounds on the estimation errors, see for instance \cite{MilVic:91}, and \cite{Hjalmarsson:05} for pointers to more recent developments. Set-membership identification may be embedded within the general framework of worst-case information-based complexity (IBC), see \cite{TrWaWo:88} and \cite{TraWer:98}, so that various systems and control problems, such as time-series analysis, filtering and $\cH_\infty$ identification can be addressed \cite{MilTem:85,Tempo:88,HeJaNe:91,SanSzn:98,GaViZa:00}. In this setting, the noise is a deterministic variable bounded within a set of given radius. The objective is to derive optimal
algorithms which minimize (with respect to the noise) the maximal distance between the true-but-unknown system parameters and their estimates. 
The main drawback of this deterministic approach is that in many instances the resulting worst-case bounds could be too conservative, and therefore of limited use, in particular when the ultimate objective is to use system identification in the context of closed-loop control.

The worst-case setting is based on the ``concern" that the noise may be very malicious. The computed bounds are certainly more pessimistic than the stochastic ones, but the idea is to guard against the worst-case scenario, even though it is unlikely to occur. 
These observations lead us to discuss the {\em rapprochement} viewpoint, see \cite{NinGoo:95,HaHoSc:99,ReGaLj:02,CaCaGa:09,GBCSA:03}, which has the following starting point: the measurement noise is confined within a given set (and therefore it falls under the framework of the worst-case setting), but it is also a random variable with given probability distribution (so that statistical information is used). A simple example is uniformly distributed noise with a supporting set which is that adopted by the worst-case methods. We recall that the rapprochement approach has been extensively studied in the context of control design in the presence of uncertainty, see \cite{TeCaDa:05,CalDab:06,CaDaTe:11}.
This research provides a  methodology for deriving controllers guaranteeing the desired performance specifications with high level of probability. 

The focal point of this paper is to address the rapprochement between soft and hard bounds in a rigorous fashion, with the goal to derive useful computational tools for linear estimation problems, see  \cite{DaSzTe:12a,DaSzTe:12b} for preliminary results. To this end, we adopt the general abstract formulation of IBC which allows to study under the same framework the two main approaches to system identification discussed so far, and to obtain new results for the probabilistic framework. In particular, the objective is to compute (by means of randomized and deterministic algorithms) the so-called {\em probabilistic radius of information}.
We remark that, contrary to the statistical setting which mainly concentrates on asymptotic results, the probabilistic radius introduced in this paper provides a quantification of the estimation error which is based on a finite number of observations. In this sense, this approach has close relations
with the works based on statistical learning theory proposed in \cite{KarVid:02,Weyer:00,VidKar:07}, and with the approach in \cite{CamWey:05,CamWey:10}, where 
distribution-free non-asymptotic confidence sets for the estimates are derived. Furthermore, the paper is also related to the work \cite{TjaGar:02}, where a probability density function over the consistency set is considered.

We now provide a preview of the structure and main results of the paper. Section \ref{sec-IBC} presents an introduction to information-based complexity and an example showing how system parameter identification and prediction may be formulated in the general IBC framework.
Section \ref{sec-prob} introduces the probabilistic setting and shows a tutorial example regarding estimation of the parameters of a second order model corrupted by additive noise. The example in continued in other sections of the paper for illustrative purposes.
In this context, the idea is to ``discard" sets of (probabilistic) measure at most $\epsilon$ from the consistency set. That is, the objective is to decrease significantly the worst-case radius, thus obtaining a new error which represents the probabilistic radius of information, at the expense of a probabilistic risk $\epsilon$. 
This approach may be very useful, for example, for system identification in the presence of outliers \cite{BCTY:02}, where ``bad measurements'' may be discarded. In this section, by means of a chance-constrained approach \cite{NemSha:06}, we also show that the probabilistic radius is related to the minimization of the so-called \textit{optimal violation function} $v\ped{o}(r)$. 

Section \ref{sec-uniform} deals with uniformly distributed noise and contains the main technical results of the paper. In particular, Theorem \ref{theorem1} shows that the induced measure over the so-called consistency set is uniform. Theorem \ref{th-difference} proves crucial properties, from the computational point of view, of the optimal violation function $v_o(r)$. In particular, this result shows that $v\ped{o}(r)$ is non-increasing, and for fixed $r>0$, it can be obtained as the maximization of a specially constructed unimodal function. Hence, it may be easily computed by means of various optimization techniques which are discussed in the next section.

In Section \ref{sec-randomized} we introduce specific algorithms for computing the 
optimal violation function. First, we observe that the exact  computation of $v\ped{o}(r)$ requires the evaluation of the volume of polytopes. Since this problem is NP-Hard \cite{Khachiyan:93}, we propose to use suitable probabilistic and deterministic relaxations. More precisely, first we present a randomized algorithm based upon the classical Markov Chain Monte Carlo method \cite{Spall:03,Spall:03b}, which has  been studied in the context of  stochastic approximation methods \cite{Chen:02,KusYin:03}; see also \cite{TeCaDa:05} and 
\cite{CaDaTe:11}  for further details about randomized algorithms. Secondly, we present a deterministic relaxation of $v\ped{o}(r)$ which is based upon
the solution of a semi-definite program (SDP). The performance of both algorithms is compared using the example previously introduced. 

Section \ref{sec-normal} discusses normally distributed noise, and presents some connections with 
classical stochastic estimation. In particular, it is shown that the least-squares algorithm is ``almost optimal'' also
in the probabilistic setting discussed in this paper. For this case, we state a bound (which is essentially tight for small-variance noise) on the probabilistic radius of information, which is given in \cite{TrWaWo:88} in terms of the so-called \textit{average radius of information}. This bound depends on $\epsilon$, on the noise covariance, and on the  so-called information and solution operators. 

Finally. in Section \ref{sec-examples} we study a numerical example of a FIR system affected by uniformly distributed noise. First, we compute deterministic and randomized relaxations of the optimal violation function. Then, by means of an extensive numerical simulation, we compare  the  probabilistic optimal estimate with  classical least-squares and the worst-case optimal estimates.

\section{Information-Based Complexity for System Identification}
\label{sec-IBC}
\begin{figure}[!ht]
\centerline{
\includegraphics[width=8.5cm]{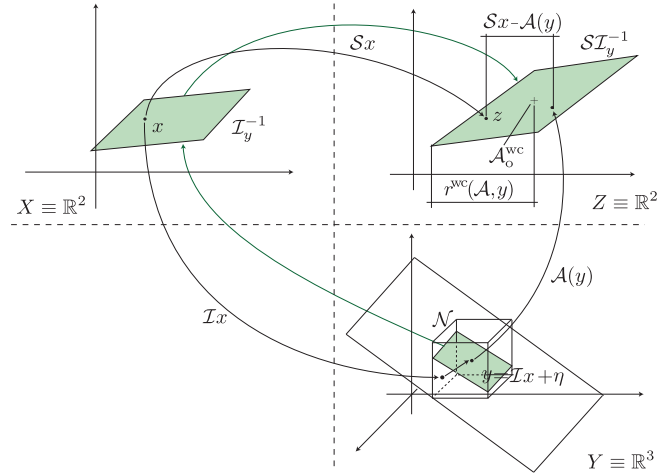}
}
\caption{Illustration of the  information-based complexity framework.\label{fig_IBC_MAIN}}
\end{figure}

This section introduces the formal definitions used in information-based complexity and an illustrative example
regarding system identification and prediction.
The relevant spaces, operators and sets
discussed next are shown in Figure~\ref{fig_IBC_MAIN}. 

Let $X$ be a linear normed $n$-dimensional space over the real field, 
which represents the set of (unknown) problem elements $x \in X$.
Define a linear operator $\cI$, called {\it  information operator}, which maps
$X$ into a linear normed $m$-dimensional space $Y$
\[ 
\cI: X \rightarrow Y. 
\]
In general, exact information about the problem element $x \in X$ is not available and only
perturbed information, or {\it data}, $y \in Y$ is given. That is, we have
\beq
\label{meas}
y = \cI x + \q
\eeq
where $\q$ represents additive \textit{noise} (or uncertainty) which may be deterministic or 
random. We assume that $\q \in \Q$, where $\Q \subseteq \Real{m}$ is a possibly unbounded
 set. Due to the presence of uncertainty $\q$, 
the problem element $x \in X$ may not be easily recovered knowing data $y \in Y$. 
Then, we introduce a  linear operator ${\cal S}$, called a {\it solution operator}, which maps $X$ into $Z$
\[
{\cal S}: X \rightarrow Z 
\] 
where $Z$ is a linear normed $s$-dimensional space over the real field, where $s \le n$. Given ${\cal S}$, our aim is to estimate an element ${\cal S} x \in Z$ 
knowing the corrupted information $y \in Y$ about the problem element $x \in X$.

An {\it algorithm} $\cA $ is a mapping (in general nonlinear) from $Y$ into $Z$, i.e.
\[
\cA : Y \rightarrow Z. 
\]
An algorithm provides an approximation $\cA (y)$ of ${\cal S} x$ using the available information $y \in Y$
of $x \in X$. The outcome of such an algorithm is called an {\it estimate} $z= \cA (y)$.
 
We now introduce a set which plays a key role in the subsequent definitions of radius of information and optimal algorithm. Given data $y \in Y$, we define the {\it consistency set} as follows
\beq
\Ix  \doteq \left\{ x \in X \,:\,  y = \cI x + \q \text{ for some } \q \in \Q
\right\}
\label{set}
\eeq
which represents the set of all problem elements $x \in X$ compatible with (i.e. not invalidated by)  $\cI x$, uncertainty $\q$ and bounding set $\Q$.
For the sake of simplicity, we assume that the three sets $X,Y,Z$ are equipped with the same $\ell_{p}$ norm. Also, in the sequel we assume that the information operator $\cI$ is a one-to-one mapping, i.e.
$m\ge n$ and ${\rm rank } \ \cI =n$. Similarly, $n\ge s$ and $\cS$ is full row rank.
Moreover, we assume that the set $\Ix$ has non-empty interior.
Note that, in a system identification context, the assumption on $\cI$ and  on the consistency set $\Ix$ are necessary conditions for identifiability of the problem element $x\in X$.
Similarly, the assumption of full-rank $\cS$ is equivalent to assuming that the elements of the vector $z=\cS x$ are linearly independent (otherwise, one could always estimate a linearly independent set and use it to reconstruct the rest of the vector $z$).
We now provide an illustrative example showing the role of these operators and spaces in the context of system identification; note that the IBC theoretical setting also applies to filtering problems, see for instance \cite{Tempo:88,GaViZa:00}.

\begin{example}[System parameter identification and prediction]
\label{ex1}
Consider a parameter identification problem which has the objective to
identify a linear system from noisy measurements.
In this case, the problem elements  are represented by
the  trajectory $\xi=\xi(t,x)$ of a dynamic 
system, parameterized  by  some  unknown  parameter  vector  $x\in  X$. 
This may be represented as the following finite regression
\[
\xi(t,x)=\sum_{i=1}^{n} x_{i} \psi_{i}(t)=\Psi\tran  (t) x,
\]
with given basis functions $\psi_i(t)$, and $\Psi\tran (t) \doteq  
\left[\psi_{1}(t)\quad\cdots\quad\psi_{n}(t) \right]$. 
We suppose that $m$ noisy measurements of $\xi(t,x)$ are available for $t_1<t_2<\cdots<t_m$,
that is
\beq
\label{noisy-info}
 y=\cI x+\q= \left[\Psi(t_{1})\quad\cdots\quad\Psi(t_{m}) \right] \tran x+\q.
\eeq
In this context, one usually assumes unknown but bounded errors, such that $|\q_{i}|\le \rho$, $i=1,\ldots,m$,
that is $\Q=
\left\{
\q\,:\, \|\q\|\le \rho
\right\}$. Then, the aim is to obtain a parameter estimate using the data $y$.
Hence, the solution operator is given by the identity, 
\[
\cS x=x
\] 
and $Z\equiv X$.
The consistency set is sometimes referred to as feasible parameters set, and is 
given as follows
\beq
\Ix  = \left\{ x \in X: \|y- \left[\Psi(t_{1})\,\cdots\,\Psi(t_{m}) \right] \tran  x\|_{\infty}\le \rho
\right\}.
\label{set23}
\eeq
For the case of time series prediction, we are interested on predicting $s$ future values of the function $\xi(t,x)$ based on $m$ past measurements, and  the solution operator takes the form
\beas
z&=&\cS x=\left\{\xi(t_{m+1},x),\quad\ldots\quad,\xi(t_{m+s},x)\right\}\\
&=& \left[\Psi(t_{m+1})\quad\cdots\quad\Psi(t_{m+s}) \right]\tran x.
\eeas
\end{example}
\vskip .1in

Next, we define approximation errors and optimal algorithms when $\q$ is deterministic or random. First, we
briefly summarize the deterministic case which has been deeply analyzed in the literature, see e.g.\ \cite{MilTem:85}. The definitions concerning the probabilistic case are new in this context, and are introduced in Section \ref{sec-prob}.

\subsection{Worst-Case Setting}

Given  data $y \in Y$, we define the worst-case error $r\apex{wc}(\cA ,y)$ of the algorithm $\cA $
as
\begin{equation}
r\apex{wc} (\cA ,y) \doteq \max_{x \in\Ix } \| {\cal S} x - \cA  (y) \|.
\label{rwc}
\end{equation}
This error is based on the available information $y \in Y$ about the
problem element $x \in X$ and it measures the approximation error between $Sx$ and $A(y)$. An algorithm $\cA \apex{wc}\ped{o}$ is called  {\it worst-case optimal} if it minimizes $r\apex{wc} (\cA ,y)$ for any $y \in Y$. That is, given data $y \in Y$, we have 
\beq
\label{rwc_opt}
r\apex{wc} \ped{o}(y) \doteq r\apex{wc} (\cA \apex{wc}\ped{o},y) \doteq
\inf_\cA  r\apex{wc} (\cA ,y).
\eeq
The minimal error $r\apex{wc}\ped{o}(y)$ is called the  {\it worst-case radius of information}.
%

This optimality criterion is meaningful  in estimation problems as it
ensures the smallest approximation error between the actual (unknown) solution ${\cal S}x$ and its
estimate $\cA (y)$ for the worst element $x \in \Ix $ for any given data $y \in Y$.
Obviously, a worst-case optimal estimate is given by $z\apex{wc}\ped{o} =
\cA \apex{wc}\ped{o}(y)$, see Figure~\ref{fig_IBC_MAIN}.

We notice that optimal algorithms map data $y$ into the $\ell_{p}$--{\it Chebychev center} of the set
${\cal S} \Ix $, where the Chebychev center $z\ped{c}(H)$ of a set $H \subseteq Z$
is defined as
\[
\max_{h \in H} \| h - z\ped{c}(H) \| \doteq  \inf_{z \in Z} \max_{h \in H} \| h - z \|\doteq r\ped{c}(H).
\]
Optimal algorithms are often called {\it  central algorithms} and 
$z\ped{c}(\cS\Ix ) = z\apex{wc}\ped{o}$ is the worst-case optimal estimate, frequently referred to as \textit{central estimate}. We remark that, in general, the Chebychev
center of a set $H \subset Z$ may not be unique (e.g.\ for $\ell_{\infty}$ norms), and not necessarily belongs to $H$, even if $H$ is convex.

\section{Probabilistic Setting with Random Uncertainty}
\label{sec-prob}
In this section, we introduce a probabilistic counterpart of the worst-case setting previously defined. That is we
define optimal algorithms $\cA \apex{pr}\ped{o}$ and the probabilistic radius $r\apex{pr} (\cA ,y,\epsilon)$ for the so-called probabilistic setting when the uncertainty $\q$ is random 
and $\epsilon\in (0,\,1)$ is a given parameter called 
{\it accuracy}.
Roughly speaking, in this setting the error of an algorithm is measured in a worst-case sense, 
but we ``discard'' a set of measure at most $\epsilon$ from the consistency set ${\cal S} \Ix $. Hence, the probabilistic
radius of information may be interpreted as the smallest radius of a ball
discarding a set whose measure is at most $\epsilon$.
 Therefore, we are decreasing the
worst-case radius of information at the expense of a probabilistic  ``risk" $\epsilon$.
In a system identification context, reducing the radius of information is clearly 
a highly desirable property. Using this probabilistic notion,  we compute a trade-off function which  shows how the radius of information decreases as a function of the parameter $\epsilon$, as described in the tutorial Example \ref{ex-MA2} and in the numerical example presented in Section \ref{sec-examples}.

Formally, in the sequel we assume that the uncertainty $\q$ is a real random
vector with given probability measure   $\muQ$ over the support set $\Q\subseteq \Real{m}$.

\begin{remark}[Induced measure over $\Ix$]
\label{rem-cond-meas}
We note that the probability measure over the set $\Q$ induces, by means of  equation (\ref{meas}),  a probability measure $\muX$ over the set $\Ix$.
This \textit{induced measure}\footnote{The induced measure $\mu_{\Q}$ is such that, for any Borel measurable set $B\subseteq X$, we have:
$\muX(B)=\mu_{\Q}(\q\in\Q\,:\, \exists x\in B\cap \Ix \text{ such that } \cI x+\q=y)$.}
is formally defined in \cite[Chapter 6]{TrWaWo:88} and it is such that points outside the consistency set $\Ix $ have measure zero, and  $\muX\left( \Ix  \right)=1$. That is, the induced measure is concentrated over $\Ix$.
We remark that Theorem \ref{theorem1} in Section \ref{sec-uniform}
studies the induced measure $\muX(\cdot)$ over the set $\Ix $ when $\q$ is uniformly distributed within $\Q$, showing that this measure is still uniform.
In turn, the induced measure $\muX$ is mapped through the linear operator $\cS$ into a measure over $\cS\Ix$, which we denote as $\muZ$. 
In Theorem \ref{theorem1} in Section \ref{sec-uniform} we show that the 
induced measure $\muZ$ is in general log-concave in the case of uniform density over $\Q$.
\erem
\end{remark}
\vskip .3in

Given  corrupted information $y \in Y$ and accuracy $\epsilon \in (0,1)$, we define the 
probabilistic error  (to level $\epsilon$) $r\apex{pr}(\cA ,y,\epsilon)$ of the algorithm $\cA $ as
\begin{equation}
r\apex{pr} (\cA ,y,\epsilon)\doteq
\inf _{\Xe \text{ such that } \muX(\Xe ) \le \epsilon
}\,
\max_{x \, \in \,
\Ix  \setminus \Xe 
}\| {\cal S} x - \cA  (y) \|
\label{rpr}
\end{equation}
where the notation $\Ix  \setminus \Xe $ indicates the set-theoretic difference between $\Ix $ and $ \Xe$.
Clearly, $r\apex{pr} (\cA ,y,\epsilon) \le r\apex{wc} (\cA ,y)$
for any algorithm $\cA $, data $y \in Y$ and accuracy level $\epsilon \in (0,1)$, which implies a 
reduction of the approximation error in a probabilistic setting.

An algorithm  $\cA \apex{pr}\ped{o}$ is called \textit{probabilistic optimal} (to level $\epsilon$)
if it minimizes the error $r\apex{pr} (\cA ,y, \epsilon)$ for any $y \in Y$ and
$\epsilon \in (0,1)$. That is, given data $y \in Y$ and accuracy level $\epsilon \in (0,1)$, we have 
\beq
\label{rpr_opt}
r\apex{pr}\ped{o}(y,\epsilon)\doteq r\apex{pr} (\cA\apex{pr}\ped{o},y,\epsilon) = 
\inf_\cA  
r\apex{pr} (\cA ,y,\epsilon) .
\eeq
The minimal error $r\apex{pr} \ped{o}(y,\epsilon)$ is called the 
{\it probabilistic radius of information} (to level $\epsilon$) and the corresponding optimal estimate is given by
\beq
\label{prob-opt-est}
{z}\apex{pr}\ped{o}(\epsilon)\doteq \cA\apex{pr}\ped{o}(y,\epsilon).
\eeq
The problem we study in the next section is the computation of 
$r\apex{pr}\ped{o} (y,\epsilon)$ and the derivation of probabilistic optimal algorithms $\cA \apex{pr}\ped{o}$. 
To this end, as in \cite{TrWaWo:88}, we reformulate equation (\ref{rpr}) in terms of a chance-constrained optimization problem \cite{NemSha:06} 
\[
\label{chance}
r\apex{pr} (\cA ,y,\epsilon) = \min \left\{r\,:\,
 v(r,\cA) \le \epsilon\right\},
\]
where  the violation function for given algorithm $\cA$ and radius $r$ is defined as
\[
\label{vr}
v(r,\cA)\doteq\muX \left\{x \in \Ix  \,:\, \| {\cal S} x - \cA  (y) \| > r 
\right\}.
\]
Then, this formulation leads immediately to
\beq
\label{chance-opt}
r\ped{o}\apex{pr} (y,\epsilon) = 
\min \left\{r\,:\,
 v\ped{o}(r) \le \epsilon\right\},
\eeq
where  the \textit{optimal violation function} for a given radius $r$ is given by
\beq
\label{vr-opt}
v\ped{o}(r) \doteq
\inf_{\cA}\muX \left\{x \in \Ix : \|\cS x - \cA  (y) \| > r 
\right\}.
\eeq

Roughly speaking, the function $v_o(r)$ describes how the risk $\epsilon$ decreases as a function of the radius~$r$. However, the computation of $v_o(r)$ is not an easy task and requires the results proved in Section \ref{sec-uniform} and the algorithms presented in Section \ref{sec-randomized}.
To illustrate the notions introduced so far, we consider the following numerical example.
The example is tutorial, and it is sufficiently simple so that all relevant
sets are two dimensional and can be easily depicted.

\begin{example}[Identification of a second order model]
\label{ex-MA2}
Our aim is to estimate the parameters of a second order FIR model
\beq
\label{ex-MA2-meas}
y_{k}=x_{1}u_{k}+x_{2}u_{k-1}+\q_{k}, \quad k=1,\ldots,m\\
\eeq
where the input $u_{k}$ is a known input sequence.
The (unknown) nominal parameters were set to $[1.25\quad 2.35]\tran$, and $m=100$
measurements were collected generating the input sequence $\{u_{k}\}$ according to a Gaussian distribution with zero mean value and unit variance, and the measurement uncertainty $\q$ as a sequence of uniformly distributed noise with $|\q_{k}|\le 0.5$.
Note that, in this case, the operator $\cS$ is the identity, and thus $X\equiv Z$
and the sets $\Ix$ and $\cS\Ix$ coincide. That is, the goal is to estimate $z_{i}=x_{i}$, $i=1,2$.

First, the optimal worst-case radius defined in (\ref{rwc_opt}) 
and the corresponding optimal solution have been computed by solving four linear programs
(corresponding to finding the tightest box containing the polytope $\cS\Ix$). The computed worst-case optimal estimate is $z\ped{o}\apex{wc}=[ 1.2499 \quad    2.3551]\tran$ and the worst-case radius is
$r\ped{o}\apex{wc}(y)=0.0352$. 

Subsequently, we fix the accuracy level $\epsilon=0.1$, and aim at computing a probabilistic optimal radius and the corresponding optimal estimate according to definitions (\ref{rpr_opt}) and (\ref{prob-opt-est}).
By using the techniques discussed in Section \ref{sec-uniform}, we obtained  $r\ped{o}\apex{pr}(y,0.1)=0.0284$
and $z\ped{o}\apex{pr}(0.1)=[1.2480 \quad  2.3540]\tran$, which represents a $25\%$ improvement.
%
\begin{figure}[!htb]
\label{fig-es-num}
\centerline{
\includegraphics[width=7cm]{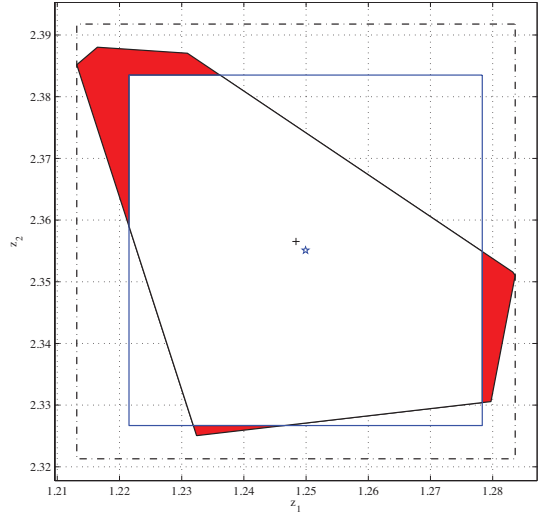}
}
\caption{Consistency set and relevant quantities for Example \ref{ex-MA2}. The worst-case optimal radius $r\ped{o}\apex{wc}(y)$ corresponds to the radius of the dash-dotted box enclosing the polytope $\cS\Ix$. Its center, denoted by a cross $+$, represents the optimal worst-case estimate $z\ped{o}\apex{wc}$.
The probabilistic optimal (to level $\epsilon=0.1$) radius  $r\ped{o}\apex{pr}(y,\epsilon)$ corresponds to the radius of the solid-line box, which is the ``optimal set'' $\cX_{\epsilon}$ that, according to definition (\ref{rpr}),
discards a set of measure $\epsilon$ from  $\Ix$. The discarded set $\Ix  \setminus \Xe$
is represented by the dark (red) area. The center of this box, denoted by a star $\star$, represents the optimal probabilistic estimate $z\ped{o}\apex{pr}(\epsilon)$. }
\end{figure}

\end{example}

\section{Random Uncertainty Uniformly Distributed}
\label{sec-uniform}

In this section, which contains the main technical results of the paper, we study the case when $\q$ is uniformly distributed over a norm bounded set, and we prove how in this case the computation of the optimal violation function, and thus of the probabilistic optimal estimate, can be formulated as a concave maximization problem. Formally, for a set $A$, the uniform density over $A$ is defined as
\[
\pU{A}(x) \doteq 
\left\{
\label{uniform}
\begin{array}{cl} \displaystyle 1/\vol{A} & \mbox{ if } x \in A; \\
               0        & \mbox{ otherwise} 
\end{array}
\right.
\]
where $\vol{A}$ represents the Lebesgue measure (volume) of the set $A$, see \cite{Halmos:50} for details on Lebesgue measures and integration.  
Note that the uniform density $\pU{A}$  generates a uniform Lebesgue measure $\lambda_{A}$ on $A$,   
such that, for any Borel measurable set $B$, $\lambda_{A}(B)=\vol{B\cap A}/\vol{A}$.

\begin{assumption}[Uniform noise over $\cB(r)$]
\label{ass-uniform}
We assume that  $\q$ is uniformly distributed over the $\ell_{p}$ norm-ball $\cB(r)=\left\{
\q\,:\, \|\q\|_p\le r \right\}$; that is, $\Q=\cB(r)$ and $\mu_{\Q}=\lambda_{\Q}$. 
\end{assumption}

First,  we address a preliminary technical question:
\textit{If  $\muQ$ is the uniform measure  over $\Q$, what
is the induced  measure $\muX$ over the set  $\Ix $ defined in equation (\ref{set})?} 
The next result shows that this distribution is indeed still uniform under the 
mild assumption of compactness of~$\Q$.

\begin{theorem}[Measures over $\Ix$ and $\cS\Ix$]
\label{theorem1}
Let $\Q$ be a compact set, and let $\q\sim\cU_{\Q}$, then, for any $y \in Y$ it holds:
\ben
\item[(i)]
The induced measure $\muX$ is uniform over $\Ix$, that is 
$\displaystyle
\muX \equiv \muU{\Ix}$;
\item[(ii)]  The induced measure $\muZ$ over $\cS\Ix$ is log-concave.
Moreover, if $\cS\in\Real{n,n}$, then this measure is uniform, that is 
$\displaystyle
\muZ \equiv \muU{\cS\Ix}.
$
\een
\end{theorem}
The proof of this theorem is reported in Appendix \ref{app-theorem1}.

\begin{remark}[Log-concave measures and Brunn-Minkowski inequality]
Statement (ii) of the  theorem proves that the induced measure on $\cS\Ix$ is log-concave.
We recall that a measure $\mu(\cdot)$ is log-concave  if, for any compact sets $A$, $B$ and $\alpha\in[0,\,1]$, it holds
\[
\mu(\alpha A +(1-\alpha) B) \ge \mu(A)^{\alpha} \mu(B)^{1-\alpha}
\]
where $\alpha A + (1-\alpha) B$ denotes the Minkowski sum\footnote{
The Minkowski sum of two sets $A$ and $B$ is obtained adding every element of $A$ to every element of $B$, i.e.\ 
$A+B=\left\{a+b\,:\, a\in A, b\in B \right\}$.}
of the two sets $\alpha A$ and $ (1-\alpha) B$. 
Note that the Brunn-Minkowski inequality \cite{Schneider:93} asserts that the uniform measure over convex sets is log-concave.
Furthermore, any Gaussian measure is log-concave.
\erem
\end{remark}
\vskip .2in

We now introduce an assumption regarding the solution operator $\cS$.

\begin{assumption}[Regularized solution operator]
\label{ass-regularized}
In the sequel, we assume that the solution operator is regularized, so that $\cS=\left[\tS\; 0_{s,n-s}\right]$,
with $\tS\in\Real{s,s}$. 
\end{assumption}

\begin{remark}[On Assumption \ref{ass-regularized}]
Note that the assumption is made without loss of generality. Indeed, for any full row rank $\cS\in\Real{s,n}$, we  introduce
the change of variables $T=\left[T_{1}\; T_{2}\right]$, where $T_{1}$ is an orthonormal basis of the column space of 
$\cS\tran $ and $T_{2}$ is an orthonormal basis of the null space
 of 
$\cS$. 
Then, $T$ is orthogonal by definition, and it follows
\beas
z&=&\cS x
=\cS T T\tran  x
=\cS 
\left[
T_{1}\; T_{2}
\right]
 T\tran  x\\
&=& 
\left[
\cS T_{1}\; \cS T_{2}
\right] T\tran  x
=\left[
\tS\; 0_{s,n-s}
\right] \tilde x = \widetilde{\cS} \tilde x,
\eeas
where we introduced the new problem element $\tilde x \doteq T\tran  x$ and the new solution operator $\widetilde{\cS}\doteq \cS T$. Note that, with this change of variables,  equation (\ref{meas})  is rewritten as
$
y=\widetilde \cI \tilde x +\q,
$ 
by introducing the transformed information operator $\widetilde\cI \doteq \cI T$. We observe that any algorithm $\cA$, being a mapping from $Y$ to $Z$, is invariant to this change of variable.
It is immediate to conclude that the new problem defined in the variable $\tilde x$ and the operators $\widetilde \cI$ and $\widetilde \cS$ satisfies Assumption \ref{ass-regularized}.
\erem
\end{remark}
\vskip .3in

Instrumental to the next developments, we introduce the
cylinder in the element space $X$,  with given ``center'' $z\ped{c}\in Z$ and radius $r$, as follows
\bea
\label{cylinder}
\Cyl(z\ped{c},r)&\doteq& \left\{x\in\Real{n} \,:\, 
\|\cS x-z\ped{c}\|\le r
\right\}\\
& =& \cS^{-1}(\cB(z\ped{c},r)) \subset X, \nonumber
\eea
that is, $\Cyl(z\ped{c},r)$ is the inverse image (pre-image) under the solution operator $\cS$ of 
the $\ell_{p}$ norm-ball $\cB(z_{c},r)\doteq\left\{z\,:\, \|z-z_{c}\|_p\le r\right\}$.
Moreover, due to Assumption \ref{ass-regularized}, the cylinder $\Cyl(z\ped{c},r)$ is parallel to the coordinate axes, that is any element $x$ of the cylinder can be written as 
\beas
\lefteqn{x\in\Cyl(z\ped{c},r) \Leftrightarrow }\\
x=&\left[\ba{c} \tS^{-1}\zeta\\ \xi \ea\right],  \zeta\in \cB(z\ped{c},r)\subset \Real{s},\, \xi\in\Real{n-s}. 
\eeas
Hence, for the case $s<n$, the cylinder is unbounded, while for $s=n$ it is simply a linear transformation through $\cS^{-1}$ of an $\ell_{p}$ norm-ball.
Next, for given center $z\ped{c}\in Z$ and radius $r>0$, we define the intersection set between  the cylinder $\Cyl(z\ped{c},r)$ and the consistency set~$\Ix$  
\beq
\label{Phi}
\Phi(z\ped{c},r) \doteq  \Ix\,  \cap\, \Cyl(z\ped{c},r)\subset X
\eeq
and its volume
\beq
\label{phi}
\phi(z\ped{c},r) \doteq  \vol{\Phi(z\ped{c},r)}.
\eeq
Finally, we define the set $\cH(r)$ of all centers $z\ped{c} \in \Real{s}$ for which the intersection set  
$\Phi(z\ped{c},r)$
is non-empty, i.e.
\beq
\label{centers}
\cH(r)\doteq\left\{
z\ped{c}\in\Real{s}\,:\, \Phi(z\ped{c},r) \not = \emptyset
\right\}.
\eeq
Note that, even if the cylinder $\Cyl(z\ped{c},r)$ is in general unbounded, the set $\Phi(z\ped{c},r)$ is bounded whenever $z\ped{c}\in\cH(r)$, since  $\Ix$ is bounded for uniform distributions.

We are now ready to state the main theorem of this section, that provides useful 
properties from the computational point of view of the optimal violation function defined in (\ref{vr-opt}).

\begin{theorem}
\label{th-difference}
Under Assumptions \ref{ass-uniform} and \ref{ass-regularized}, the following statements hold
\ben
\item[(i)]
For given $r>0$, the optimal violation function $v\ped{o}(r)$ is given by  
\beq
\label{Vdiff}
v\ped{o}(r)=
1 -  \frac{\vstar(r)}{\vol{\Ix}},
\eeq
where $\vstar(r)$ is the solution of the optimization problem
\beq
\label{Vstar}
\vstar(r) \doteq \max_{z\ped{c}\in \cH(r)}  \phi(z\ped{c},r)
\eeq
with  $\phi(z\ped{c},r)$ and $\cH(r)$  defined in (\ref{phi}) and (\ref{centers}), respectively;
\item[(ii)]
For given $r>0$, the function $\phi(z\ped{c},r)$ is quasi-concave\footnote{A  function $f$ defined on a convex set
$A\in\Real{n}$ is quasi-concave if
$f(\alpha x + (1-\alpha)y)\ge\min(f(x),f(y))$
holds for any  $x,y\in A$ and $\alpha\in (0,\,1)$.} for $z\ped{c}\in\cH(r)$, and the set $\cH(r)$ is convex;

\item[(iii)]
The function 
$v\ped{o}(r)$ is right-continuous and non-increasing for $r>0$.
\een
\end{theorem}
The proof of this result is reported in Appendix \ref{app-th-difference}.
\vskip .1in

\begin{remark}[Unimodality of the function $\phi(z\ped{c},r)$]
Point (ii) in Theorem \ref{th-difference} is crucial from the computational viewpoint. Indeed, as remarked for instance in \cite{BoyVan:04}, a  quasi-concave function cannot have local maxima. Roughly speaking, this means that the function $\phi(\cdot,r)$ is unimodal, and therefore any local maximal solution of problem (\ref{Vstar}) is also a global maximum.
Note that from the Brunn-Minkowski inequality it follows that, if there are multiple points $z\ped{o}^{(i)}$ where $\phi(\cdot)$ achieves its global maximum, then the sets $\Phi(z\ped{o}^{(i)},r)$ are all homothetic,
see \cite{Schneider:93}. 

\begin{figure*}[!ht]
\label{fig-es1-3D}
\centerline{
\includegraphics[width=7cm]{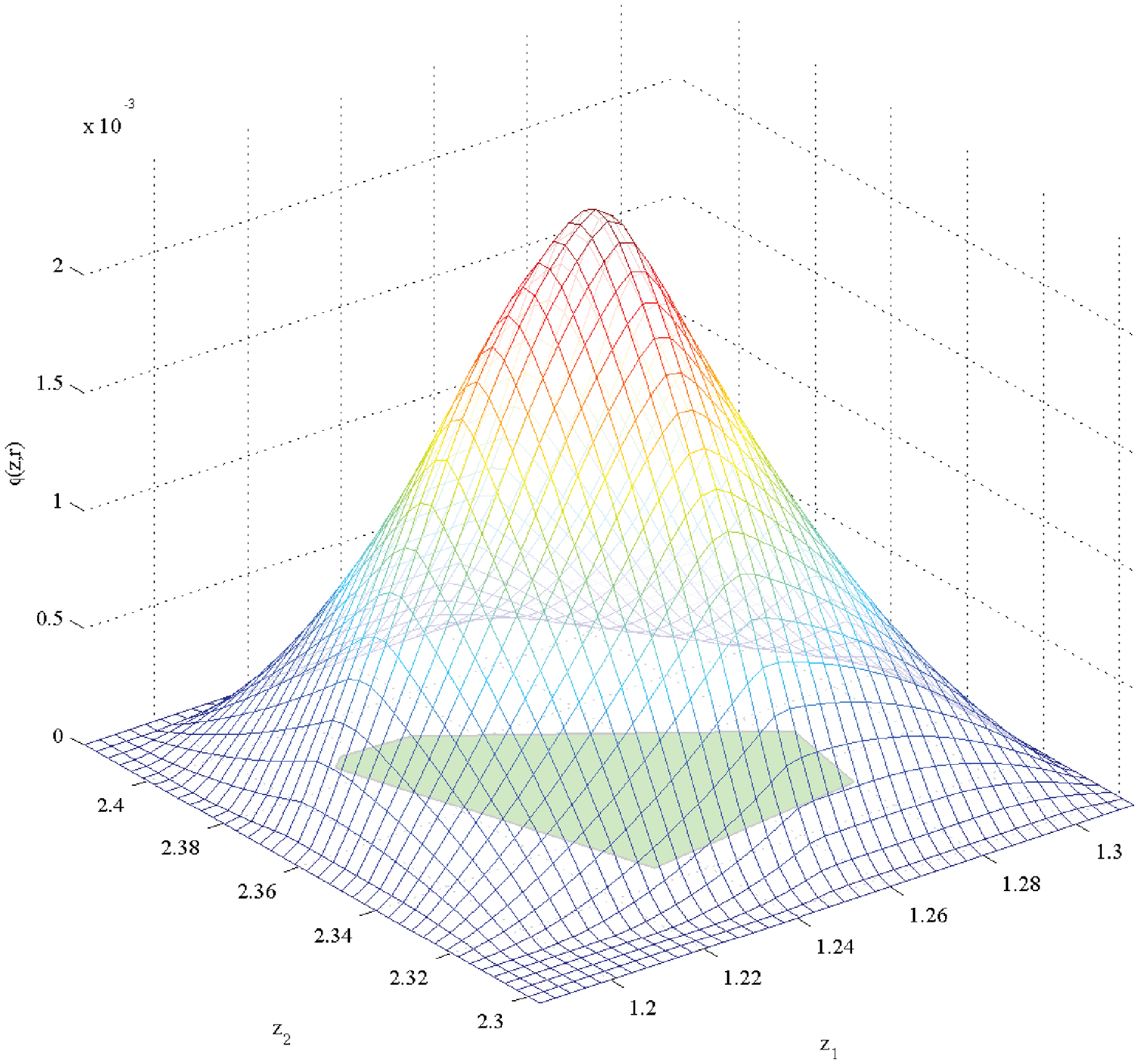}\,
\includegraphics[width=7cm]{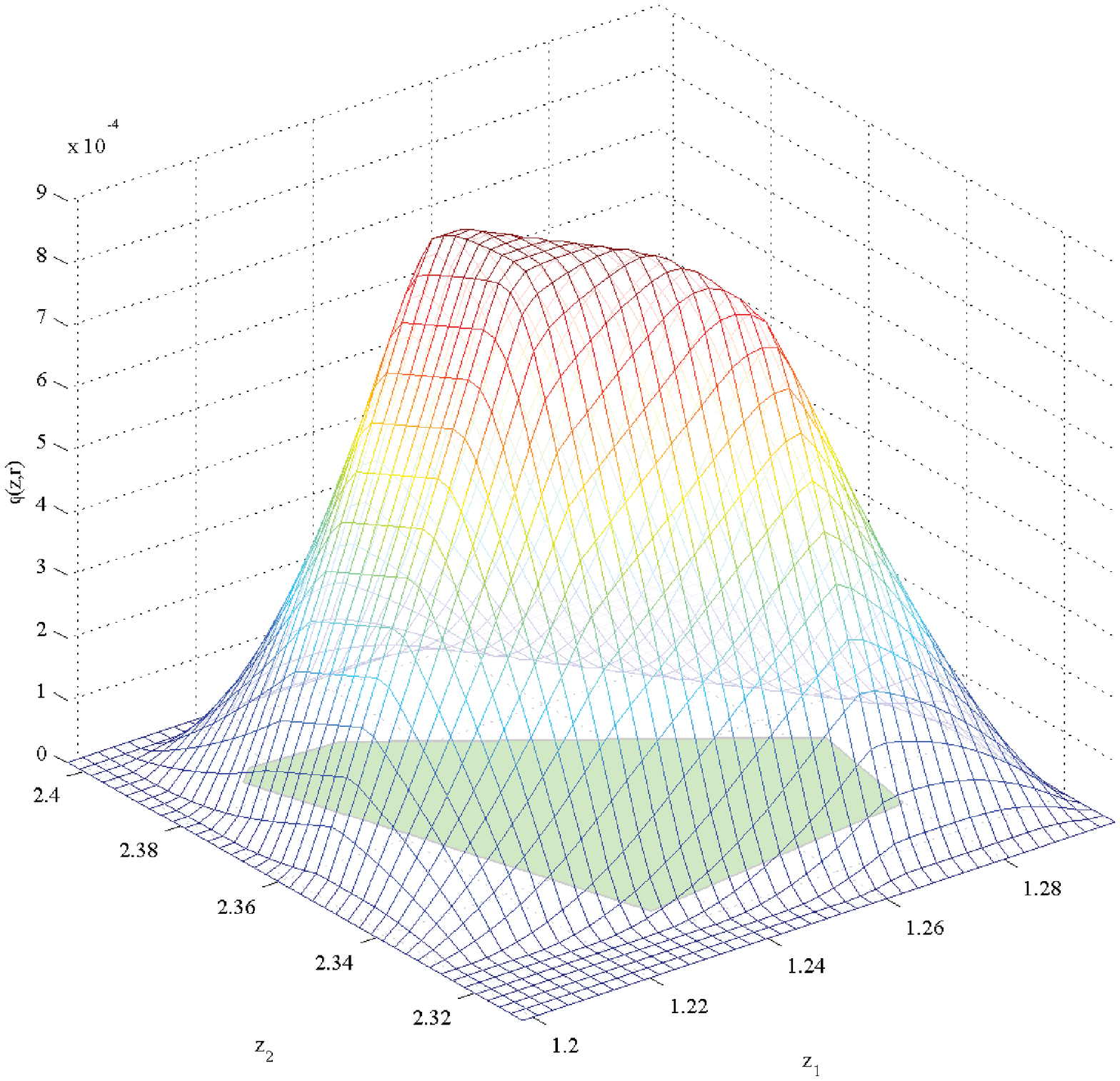}
}
\caption{The function $\phi(z\ped{c},r)$
for the tutorial problem considered in Example \ref{ex-MA2}, for  $r=0.0284$ (a) and $r=0.0150$ (b).}
\end{figure*}

These facts are illustrated in Figure \ref{fig-es1-3D}, where we plot the function $\phi(z\ped{c},r)$
for the tutorial problem considered in Example \ref{ex-MA2}, for two different values of $r$.
In the figure on the left, the two sets $\Ix$ and $\Cyl(z\ped{c},r)$ intersect for all considered values of $z\ped{c}$,  the function is unimodal, and clearly presents a unique global maximum.
In the figure on the right, the radius $r$ is smaller, and there are values of $z\ped{c}$ for which 
$\Cyl(z\ped{c},r)$ is completely contained in $\Ix$, thus leading to the ``flat'' region on the top. However, note that this is the only flat region, so that the function is ``well-behaved'' from an optimization viewpoint.
\erem
\end{remark}
\vskip .3in

\begin{remark}[Probabilistic radius and probabilistic optimal estimate]
 Theorem \ref{th-difference} provides a way of computing the optimal probabilistic radius of information $r\ped{o}\apex{pr}(y,\epsilon)$.
Indeed, the probabilistic radius of information (to level $\epsilon$)  is  given by the solution of the following one-dimensional inverse problem
\beq
\label{inverse}
r\apex{pr} \ped{o}(y,\epsilon) =  \min  \left\{ r \,:\,
v\ped{o}(r)\le \epsilon \right\}.
\eeq
Note that point (iii) in Theorem \ref{th-difference} guarantees that such solution always exists for 
$\epsilon\in(0,\,1)$, and it is unique.
The corresponding optimal estimate is then given by 
\[
z\ped{o}\apex{pr}(\epsilon)= \cA\apex{pr}\ped{o}(y,\epsilon)=z\ped{o}(r\apex{pr} \ped{o}(y,\epsilon)),
\]
where we denoted by $z\ped{o}(r)$ a solution of the optimization problem (\ref{Vstar}).

\begin{figure}[!ht]
\label{fig-es-boxes}
\includegraphics[width=8cm]{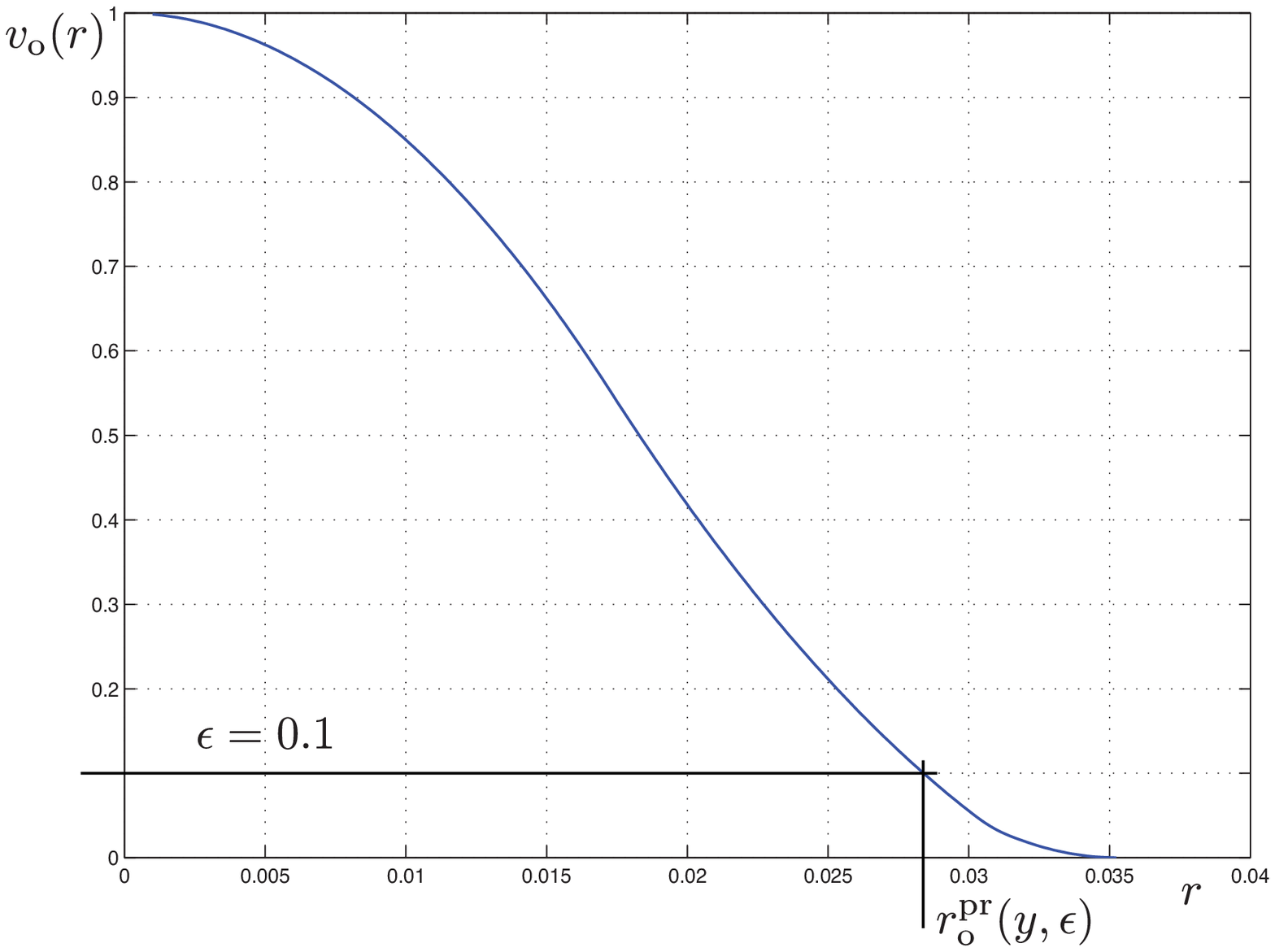}
\quad
\includegraphics[width=8cm]{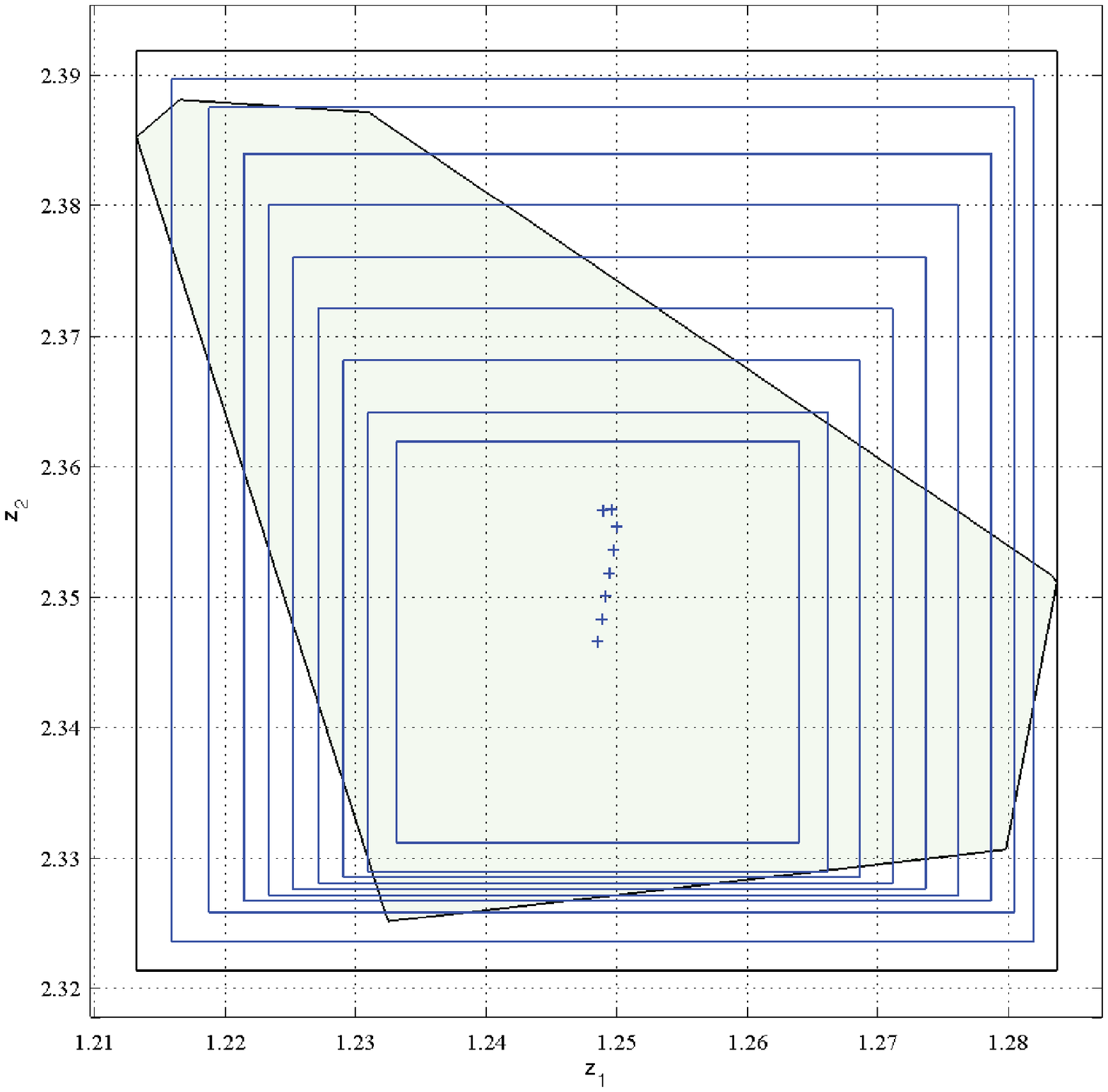}
\caption{(a) Plot of the function $v\ped{o}(r)$ for Example \ref{ex-MA2}, and computation of 
the optimal probabilistic radius for $\epsilon=0.1$. (b) Plot of the ``optimal'' $\ell_{\infty}$ balls for different values
of $r$. The crosses denote the corresponding optimal estimates $r\ped{o}\apex{pr}(y,\epsilon)$.}
\end{figure}

To illustrate, continuing Example \ref{ex-MA2}, we plot in Figure \ref{fig-es-boxes}(a) the  function $v\ped{o}(r)$ for $r\in(0,\,r\ped{o}\apex{wc}]$. We see that $v\ped{o}(r)$ is indeed non-increasing 	(actually, it is strictly decreasing), and hence the inverse problem (\ref{inverse}) has clearly a unique solution for any $\epsilon\in(0,\,1)$.
\erem
\end{remark}
\vskip .3in

Theorem \ref{th-difference} shows that the problem we are considering is indeed a well-posed one,
since it has a unique solution (even though not a unique minimizer in general). However, its solution requires the computation of the volume of the intersection set $\Phi(z\ped{c},r)$, which is in general a very hard task. A notable exception in which the probabilistic optimal estimate is immediately computed for $\q$ uniformly distributed in $\Q$ is the special case when the consistency set $\Ix$ is 
centrally symmetric\footnote{
A set $H$ is said to be \textit{centrally symmetric} with center $\bar{x}$ if $x\in H$ implies that its reflection with respect to $\bar x$ also belongs to $H$, i.e.\ $(2\bar x-x)\in H$.} with center $\bar x$. Indeed, in this case it can be seen that
 $\cS\Ix$ is also centrally symmetric around $\bar z=\cS \bar x$, and so is the density $\muZ$. Hence, the optimal probabilistic estimate coincides with the center $\bar z$, since it follows from symmetry that the probability measure of the intersection of $\cS\Ix$ with an $\ell_{p}$ norm-ball is maximized when the two sets are concentric. Moreover, this estimate coincides with the classical worst-case (central) estimate, which in turn coincides with the classical least squares estimates.

\begin{remark}[Weighted $\ell_{2}$ norms] 
\label{rem-ell2}
Note that the requirement of $\Ix$ being centrally symmetric is quite demanding in general, but holds naturally when (weighted) $\ell_2$ norms are considered, that is when  $\q$  is uniformly distributed in a the ball
\[
\Q = \left\{\q: \|\q\|_{W} \le \rho \right\}, \,\|\xi\|_{W}\doteq \sqrt{\xi\tran  W \xi}, W =W\tran\succ 0, 
\]
with $W \succ 0$ meaning positive definite, and $\mu_\q(\Q) = \lambda_{\Q}$.
This framework has been also considered in the classical set-membership literature, see for instance \cite{KMTV:86}, and it is well-known that in this case the set  $\Ix $ is the ellipsoid 
\bea
\label{ellip}
\Ix&=&\left\{x\in X : \|\cI x -y\|_{W}\le \rho\right\}
 \\
&=&\left\{x\in X : (x-x\apex{ls})\tran (\cI\tran W\cI)  (x-x\apex{ls})\le \rho^{2}\right\},\nonumber
\eea
centered around the (weighted) least-squares optimal parameter estimate
\[
x\apex{ls}\doteq \left(\cI\tran  W \cI\right)^{-1} \cI\tran  W  y.
\]
Hence, it follows from symmetry that,  for any $\epsilon\in (0,1)$, the probabilistic optimal estimate 
to level $\epsilon$ is given by
\[
z\ped{o}\apex{pr}(\epsilon)=\cA \apex{ls}(y)=
{\cal S} x\apex{ls}(y).
\]
However, we are not aware of any closed-form equation for the corresponding 
probabilistic optimal radius $r\ped{o}\apex{pr}(y,\epsilon)$,
see Section \ref{sec-normal} for further comments.
\erem
\end{remark}

\begin{remark}[Connections with worst-case and MLE estimates]
Since the paper considers a setup which is somehow in-between classical statistical estimation and set-membership estimation, it is of interest to discuss the differences and analogies between the various approaches.
The advantages with respect to the worst-case based set-membership approach, in terms of conservatism reduction, should be evident from the discussion so far, and will be further analyzed in the numerical example of Section \ref{sec-examples}.

To better clarify the connections with 
classical stochastic maximum-likelihood estimation (MLE), note that  in  \cite{TemWas:88} it is shown that, for the case of uniform noise, the MLE estimates are not unique, and any element of $\cS\Ix$ is an MLE estimate. Hence, any approach returning estimates belonging to the consistency set is optimal in this sense. In the IBC literature, estimates with the property of belonging to the consistency set are called \textit{interpolatory}, see eg.\ \cite{TrWaWo:88} for a formal definition. Interpolatory estimates enjoy interesting properties: for instance, it is easy to show that they are almost worst-case optimal (within a factor of 2).
In particular, it can be shown, using results from convex analysis \cite{Amir:86}, that in the case of uniform noise bounded in the $\ell_{2}$ norm, both the central estimate obtained in the set-membership approach and the probabilistic optimal estimate are indeed interpolatory. Hence, in this case, our approach can be seen as a tool for selecting an optimal MLE solution.

The situation is more complicated for $\ell_{1}$ or $\ell_{\infty}$ norms, because in this case the central estimate is not always interpolatory\footnote{This is a consequence of the fact that the Chebychev center of a convex set may lie outside of the set for non-Euclidean norms. Consider for instance the tetrahedron formed by the convex hull of the points $[1\, 1\, 1]\tran$, $[-1\, 1\, 1]\tran$, $[1\, -1\, 1]\tran$, $[1\, 1\, -1]\tran$. It is easy to check that the origin is the (unique) Chebychev center in the $\ell_{\infty}$ norm of the set, and it lies outside of it. Note that the fact that the central estimate can be non-interpolatory is not always clearly evidenced in the set-membership literature.}.
Similarly, the probabilistic optimal estimate defined in (\ref{rpr_opt}) is not necessarily interpolatory. Furthermore, we note that in this case also the classical least-squares estimate may lie outside  the consistency set $\cS\Ix$, hence it is not MLE. For instance, in Example \ref{ex-MA2} the least-squares estimate  can be immediately computed as $x\apex{ls}=[1.2873\quad  2.3190]\tran$ and it is indeed not interpolatory. 

An interesting approach, in the case of $\ell_{1}$ or $\ell_{\infty}$ norms, could be indeed to 
consider a conditional probabilistic-optimal estimate,which requires looking for the best interpolatory estimate minimizing the probabilistic radius (\ref{rpr}). Note that, from a computational viewpoint, this is immediately obtained constraining the optimization problem (\ref{Vstar}) to $z\ped{c}\in\cS\Ix$.
\end{remark}

\section{Randomized and deterministic algorithms for optimal violation function approximation}
\label{sec-randomized}

In this section, we concentrate on the solution of the  optimization problem
defined in (\ref{Vstar}), Theorem \ref{th-difference} for fixed $r>0$. For simplicity, we restate this problem dropping the subscript  from $z\ped{c}$
\bea
\label{Pmax}
\text{\Pmax: } &&
\max_{z\in\cH(r)} 
\phi(z,r),\\
&& \quad\phi(z,r) = \vol{ 
 \Ix\,  \cap\, \Cyl(z,r) } .
\nonumber 
\eea

First, note that this problem is computationally very hard in general. For instance, for $\ell_{1}$ or $\ell_{\infty}$ norms, the consistency set  $\Ix$ is a polytope and $\Cyl(z,r)$ is a cylinder parallel to the coordinate axes whose cross-section is a polytope.
Hence, even evaluating the function $\phi(z,r)$ appearing in (\ref{Pmax})  amounts to computing the volume of a polytope, and this problem has been shown to be  NP-hard in \cite{Khachiyan:93}.

\begin{remark}[Volume oracle and oracle-polynomial-time algorithm]
\label{rem-volume}
For the case of polytopic sets,  the papers \cite{AhChRe:10,FukUno:07} study the problem \Pmax in the hypothetical setting that an oracle exists which satisfies the following property: given $r>0$ and $z\in\cH(r)$,
it returns the value of the function $\phi(z ,r)$, together with a sub-gradient of it.
In this case, in  \cite{AhChRe:10} a strongly polynomial-time
 (\textit{in the number of oracle calls}) algorithm is derived.
Note that, even if the problem is NP hard in general, one can compute the volume of a polytope in a reasonable time for considerably complex polytopes in modest (e.g.\ for $n\le 10$) dimensions, see \cite{BuEnFu:00}. In this particular case, for $\ell_{\infty}$ norms, the method proposed by \cite{FukUno:07}
may be used. 
For instance, for Example \ref{ex-MA2}, all relevant quantities have been computed exactly by employing this method.
However it should be remarked that, for larger dimensions, the curse of dimensionality makes the problem computationally intractable, and alternative methods need to be devised.
\erem
\end{remark}
\vskip .1in

In the next subsections, we develop random and deterministic relaxations of problem \Pmax which do not suffer from these computational drawbacks.
\subsection{Randomized algorithms for computing \Pmax}

In this section, we propose randomized algorithms based on a probabilistic volume oracle and a 
stochastic optimization approach for approximately solving problem \Pmax for  $\ell_{p}$ norms. 
First of all, we compute a bounded version of the cylinder $\Cyl(z,r)$. To this end, we note that bounds $x_{i}^{-}$, $x_{i}^{+}$ on the variables $x_{i}$, $i=s+1,\ldots,n$, can be obtained as the solution of the following $2(n-s)$ convex programs, 
\bea
\nonumber
&&\ba{lll}
x_{i}^{-}=&\min\, x_{i}&\\
&\text{\rm subject to} & x\in \Ix
\ea,\\
&&\nonumber
\ba{lll}
x_{i}^{+}=&\max\, x_{i} &\\
&\text{\rm subject to} & x\in \Ix
\ea,\\
&&
i=s+1,\ldots,n. 
\label{bound}
\eea
The problems above are convex, and for  $\ell_{p}$ norms can be solved for instance by (sub)gradient-based or interior point methods. In particular, problem (\ref{bound}) reduces to the solution to $2(n-s)$ linear programs for $\ell_{1}$ or $\ell_{\infty}$ norms.
Then,  under Assumption \ref{ass-regularized}, we define the cylinder
\bea
\label{cylinder2}
\overline\Cyl(z,r)&\doteq& 
\left\{x\in\Real{n} \,:\, 
\left\|
\tS \left[
\begin{smallmatrix}
x_{1}\\
\vdots\\
x_{s}
\end{smallmatrix}
\right]
-z
\right\|\le r, 
\right.\\
&&\quad
\left. x_{i}^{-}\le x_{i}\le x_{i}^{+}, i=s+1,\ldots,n
\right\}.
\nonumber
\eea
Note that the cylinder $\overline\Cyl(z,r)$ is bounded, and has volume equal to 
\beq
\label{volume-cyl}
\vol{\overline\Cyl(z,r)} \!= \! \frac{(2r)^{s}\Gamma^{s}\left(1/p+1 \right)} 
{|\det(\tS)|\Gamma \left(s/p+1 \right)}\!\!\prod_{i=s+1}^{n}(x_{i}^{+}-x_{i}^{-})\doteq \Vcyl
\eeq
where $\Gamma(\cdot)$ denotes the Gamma function.
By construction, we have that, for any $r>0$ and $z\in\cH(r)$,
$
\Phi(z,r)= \Ix\,  \cap\, \overline\Cyl(z,r).
$
Note  that  independent and identically distributed (iid) random samples inside $\overline\Cyl(z,r)$ can be easily obtained from iid uniform samples in the $\ell_{p}$-norm ball, whose generation is studied in \cite{CaDaTe:99}.
Then, a probabilistic approximation of the volume of the intersection $\Phi(z,r)$ may be computed by means of the randomized oracle presented in Algorithm \ref{prob-oracle}, which is based on the uniform generation of iid samples in $\overline\Cyl(z,r)$. 

\begin{algorithm}[H]
\caption{Probabilistic Volume Oracle}
\label{prob-oracle}
\bit
\item[1.] \underline{\sc Random Generation}\\
Generate $N$ iid uniform samples $\zeta^{(1)},\ldots,\zeta^{(N)}$ in the $s$-dimensional ball $\cB(z,r)$
\bit
\item For $i=1$ to $N$
\bit
\item[-] Generate $s$ iid scalars $\gamma_{j} $ according to the unilateral Gamma density 
\beq
\label{gamma}
G_{a,b}(x) =
 \frac{1}{\Gamma(a)b^a} x^{a-1} \mathrm{e}^{-x/b}, \quad x \ge 0,
\eeq
with parameters $a=1/p,b=1$
\item[-] Construct the vector $\eta\in{\mathbb R}^{n}$ of com\-po\-nents $\eta_j=s_j \gamma_j^{1/p}$, where $s_j$ are iid random signs
\item[-] Let $\zeta^{(i)}=z+r\, w^{1/n} \,\frac{\eta}{\|\eta\|_p}$ where $w$ is uniform in $[0,1]$
\eit
 End for
\eit
\item[]
Generate $N$ iid uniform samples $\xi^{(1)},\ldots,\xi^{(N)}$
\bit
\item For $i=1$ to $N$
\bit
\item[-] Generate $\xi_{j}^{(i)}$ uniformly in the interval $[x_{s+j}^{-},\,x_{s+j}^{+}]$, $j=1,\ldots,n-s$
\eit
 End for
\eit
\item[]
Construct the random samples in $\overline\Cyl(z,r)$ as follows
\[
\chi^{(i)}=
\begin{bmatrix}
\tS^{-1}\zeta^{(i)}\\
\xi^{(i)}
\end{bmatrix}, \quad i=1,\ldots,N
\]
\item[2.] \underline{\sc Consistency Test}
\bit
\item Compute the number of samples inside $\Ix$ as follows
\[
N_{g}=\sum_{i=1}^{N} \ind{\| \cI\, \chi^{(i)}-y\|\le\rho}
\]
where $\ind{\cdot}$ denotes the indicator function, which is equal to one if the argument  is true, and it is zero otherwise. 
\eit
\item[3.] \underline{\sc Probabilistic Oracle} Return an approximation of the volume $\phi(z,r)$ as follows
 \[
\widehat\phi_{N}(z,r) = 
\frac{N_{g}} {N} 
\Vcyl
\]
where $\Vcyl$ is defined in (\ref{volume-cyl}).
\eit
\end{algorithm}

Note that the expected value of the random variable $\widehat\phi_{N}(z,r)$ with respect to the samples $\chi^{(1)},\ldots,\chi^{(N)}\in\overline\Cyl(z,r)$ is exactly the volume function $\phi(z,r)$ appearing in \Pmax that is
\[
\Eval{\widehat\phi_{N}(z,r)} = \phi(z,r). 
\]
This immediately follows from linearity of the expected value
\beas
\Eval{\widehat\phi_{N}(z,r)} &=&\Eval{\frac{1}{N}\sum_{i=1}^{N} \ind{\chi^{(i)}\in\Ix} \Vcyl}\\
&=&\frac{1}{N} \sum_{i=1}^{N} \Eval{\ind{\chi^{(i)}\in\Ix}} \Vcyl.
\eeas
Then, we have
\beas
\Eval{\ind{\chi^{(i)}\in\Ix}}&=& \Prob{\chi^{(i)}\in\Ix}  \\
&=& \vol{\Phi (z,r)} /  \vol{ \bar \Cyl(z,r)}\\
&=&\phi(z,r) / \Vcyl.
\eeas
Hence, we reformulate the problem  \Pmax as the following  stochastic optimization 
problem 
\[
\max_{z\in\cH(r)} \Eval{\widehat\phi_{N} (z,r)}.
\]
This problem is classical and different stochastic approximation algorithms have been proposed, see for instance \cite{KusYin:03, Shapiro:03} and references therein.
In particular,  in this paper, we use the SPSA (simultaneous perturbations stochastic approximation) algorithm,
first proposed in \cite{Spall:92}, and further discussed in \cite{Spall:03}. Convergence results under different 
conditions are detailed in the literature, see in particular the paper \cite{HeFuMa:03} which applies to 
non-differentiable functions. 

\begin{remark}[Scenario-based algorithms]
An alternative approach based on randomized methods can be also devised employing results  on the scenario optimization method introduced in \cite{CalCam:06tac}.
In particular, exploiting the results on discarded constraints, see \cite{Calafiore:10siopt, CamGar:11},
an alternative algorithm can be constructed. The idea is as follows: (i) generate $N$ samples $\chi^{(i)}$ in 
$\Ix$ according to the induced measure $\muX$, ii)
solve the discarded-constraint random program
\bea
\label{discarded}
\min_{z,\gamma} &\gamma& \\
\mbox{ s.t. } &&
\frac{1}{L}
\sum_{ i\in I_{L}}
\I \left( \| {\cal S} \chi^{(i)} -z)\| \ge \gamma \right) \le \epsilon \nonumber
\eea
where $I_{L}$ is a set of $L$ indices constructed discarding in a prescribed way
$N-L$ indices from the set $1,2,\ldots,N$.
Then, in   \cite{Calafiore:10siopt, CamGar:11} it is shown how to choose $N$ and the discarded set $I_{L}$
to guarantee, with a prescribed level of confidence, that the result of optimization problem
(\ref{discarded}) is a good approximation of the true probabilistic radius $r\ped{o}\apex{pr}(y,\epsilon)$.
However, this approach entails many seriuos technical difficulties, such as the random sample generation in point (i)
and the optimal discarding procedure in point (ii), whose detailed analysis
goes beyond the scope of this paper.
\erem
\end{remark}

\subsection{A semi-definite programming relaxation to \Pmax}
In this section, we propose a deterministic approach to \Pmax based on a semidefinite relaxation of the problem for $\ell_{\infty}$ norms (extensions to $\ell_{1}$ and $\ell_{2}$ norms are briefly discussed in Remark \ref{rem-SDP-ell2}).
First note that, in the case of $\ell_{\infty}$ norms, $\Q$ is an hypercube of radius $\rho$ and therefore $\Ix $ is the polytope $\cP_{X}$ defined by the following linear inequalities
\bea
\Ix  &=& \left\{
x\in\Real{n} \,:\, \|\cI x-y\|_{\infty}\le \rho
\right\}\\
&=& \left\{
x\in\Real{n} \,:\, 
\left[
\begin{array}{c}
 \cI  \\ -\cI
\end{array}
\right]
x \le 
\left[
\begin{array}{c}
  \rho\uni+y  \\ \rho\uni-y
\end{array}
\right]
\right\}\doteq \cP_{X}
\nonumber
\eea
where $\uni$ is a vector of ones, $\uni=[1\;1\,\cdots\,1]\tran $. 
Since the exact computation 
of  the volume of the intersection of two polytopic sets
is in general costly and prohibitive in high dimensions, as discussed in Remark \ref{rem-volume}, we propose to maximize a suitably chosen lower bound
of this volume. This lower bound can be computed as the solution of a convex optimization problem.
The idea is to construct, for fixed $r>0$,  the maximal volume ellipsoid contained in
the intersection $\Phi(z,r)$, which requires 
to solve the optimization problem
\bea
\label{ell-opt}
\max_{z,x_{\cE},P_{\cE}}&&
\vol{\cE(x_{\cE},P_{\cE})}\\
\text{subject to} &&
\cE(x_{\cE},P_{\cE}) \subseteq \Phi(z,r),
\nonumber
\eea
where the ellipsoid of center $x_{\cE}$ and shape matrix $P_{\cE}$
is
\[
\cE(x_{\cE},P_{\cE})\doteq
\left\{
x\in\Real{n} \,:\, x=x_{\cE}+P_{\cE} w, \|w\|_{2}\le 1
\right\}.
\]
The problem of deriving the maximum volume ellipsoid inscribed in a polytope is a well-studied one, and  concave reformulations based on linear matrix inequalities (LMI) are possible, see for instance \cite{BEFB:94,BenNem:98}. 
For completeness, we report this result in the next theorem.

\begin{theorem}
\label{th-ell-opt}
Let Assumptions \ref{ass-uniform} and \ref{ass-regularized} hold. Then, for given $r>0$, a center that achieves a global optimum for problem (\ref{ell-opt}) can be computed as the solution of the following semi-definite programming (SDP) problem

\bea
\lefteqn{
z\ped{o}\apex{sdp}(r)\in\arg_{z}\,\min_{z,x_{\cE},P_{\cE}} -\log\det P_{\cE}}\nonumber\\\
&&\text{\rm subject to } P_{\cE}\succeq 0\quad \text{\rm and}\nonumber\\
&&\left[
\begin{array}{cc}
(\rho+e_{i}\tran (y-\cI x_{\cE}))I_{n}  & P_{\cE}\cI\tran e_{i} \\
\star  &    \rho+e_{i}\tran (y-\cI x_{\cE})
\end{array}
\right]\succeq 0,
\nonumber\\
 && 
i=1,\ldots,m
\label{I1}\\
&&\left[
\begin{array}{cc}
(\rho-e_{i}\tran (y-\cI x_{\cE}))I_{n}  & - P_{\cE}\cI\tran e_{i} \\
\star  &    \rho-e_{i}\tran (y- \cI x_{\cE})
\end{array}
\right]\succeq 0, \nonumber\\\
&& 
 i=1,\ldots,m \label{I2} 
\eea
\bea
&&\left[
\begin{array}{cc}
(r+\bar e_{i}\tran (z - \cS x_{\cE}))I_{n}  & P_{\cE}\cS\tran \bar e_{i} \\
\star  &    r+\bar e_{i}\tran (z - \cS x_{\cE})
\end{array}
\right]\succeq 0, 
\nonumber\\ && 
 i=1,\ldots,s \label{B1}\\
&&\left[
\begin{array}{cc}
(r-\bar e_{i}\tran (z - \cS x_{\cE}))I_{n}  & P_{\cE}\cS\tran \bar e_{i} \\
\star  &    r-\bar e_{i}\tran (z - \cS x_{\cE})
\end{array}
\right]\succeq 0, 
\nonumber\\ && 
 i=1,\ldots,s, \label{B2}
\eea
where $e_{i}$ and $\bar e_{i}$ are elements of the canonical basis of $\Real{m}$ and $\Real{s}$, 
respectively. 
Moreover, for all $r>0$, $v\ped{o}\apex{sdp}(r)\ge v\ped{o}(r)$, where we defined 
\[
v\ped{o}\apex{sdp}(r)\doteq
1-\frac{\phi\left(z\ped{o}\apex{sdp}(r),r\right)}   {\vol{\Ix}}.
\]
\end{theorem}

\proof
The theorem is immediately proved seeing that (\ref{I1}), (\ref{I2}) impose that $\cE(x_{\cE},P_{\cE})\subseteq\Ix$
while (\ref{B1}), (\ref{B2}) impose that $\cE(x_{\cE},P_{\cE})\subseteq\Cyl(z,r)$. This problem is an SDP
since the equations are linear matrix inequalities  in the variables $z,x_{\cE},P_{\cE}$,
and the cost function is convex in $P_{\cE}$.\qed

From Theorem \ref{th-ell-opt}, if follows that the SDP relaxation leads to a suboptimal violation function $v\ped{o}\apex{sdp}(r)$. 

\begin{figure}[!ht]
\label{fig-es-num}
\centerline{
\includegraphics[width=8cm]{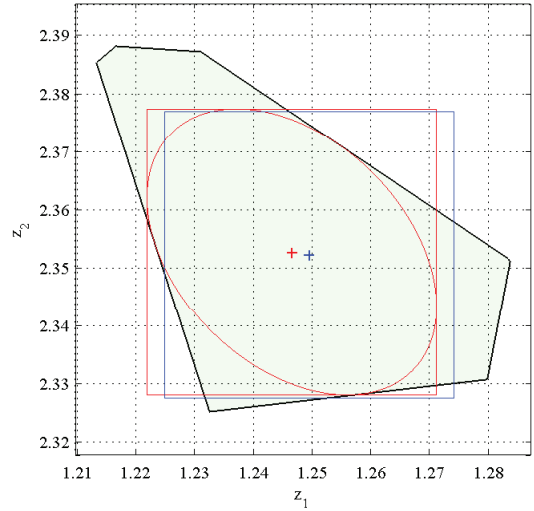}
}
\caption{Optimal $\ell_{\infty}$ ball for Example \ref{ex-MA2}, for $r=0.0284$  (in blue), 
and optimal ball computed as the solution of the SDP relaxation, with corresponding ellipse $\cE(x_{\cE},P_{\cE})$ (in red).}
\end{figure}

\begin{remark}[SDP relaxations for $\ell_{1}$ and $\ell_{2}$]
\label{rem-SDP-ell2}

An approach identical to that proposed in Theorem~\ref{th-ell-opt} can be developed for the
$\ell_{1}$ norm, considering that also in this case the sets $\Ix$ and $\Cyl(z,r)$ are a polytope and a
cylinder with polytopic basis, respectively.
Similarly, an analogous algorithm can be devised for the (weighted) $\ell_{2}$ norm. In this case, the volume of an ellipsoid contained in the intersection of $\Ix$  and $\Cyl(z,r)$ should be maximized, which are respectively the ellipsoid defined in (\ref{ellip}) and a cylinder with spherical basis.
It can be easily seen, see e.g. \cite{BEFB:94}, that this latter problem can be easily rewritten as a convex SDP optimization problem.
\erem
\end{remark}

\section{Random Uncertainty Normally Distributed and Connections with Least-Squares}
\label{sec-normal}

In this section, we concentrate on the  case  when the uncertainty $\q$  is 
normally distributed with mean value $\bar v$ and covariance matrix $\Sigma=\sigma^{2}I\succ 0$,
and the set $\Q$ coincides with $\Real{m}$. 
This permits to draw a bridge between the probabilistic setting introduced in this paper and the classical theory of statistical estimation, which is usually based on additive noise normally distributed. Indeed, it is well known, see e.g.\ \cite{KaSaHa:00,Ljung:99} that the minimum variance unbiased estimate for the linear regression model (\ref{meas}) is given by the Gauss-Markov estimate
\[
x\apex{ls}= \left(\cI\tran \Sigma^{-1} \cI\right)^{-1} \cI\tran  \Sigma^{-1}  y, 
\]
which coincides with the (weighted) least-squares estimate discussed in Remark \ref{rem-ell2},
for $W=\Sigma^{-1}$.

We first remark that this minimum variance problem falls into the average setting of IBC, see \cite{TrWaWo:88}. In particular, we recall that this setting has the objective of minimizing the expected value of the estimation error, that is, for given $y$, the optimal average radius is defined as
\begin{equation}
r\apex{av}\ped{o}(y)\doteq  \inf_\cA \left(\Eval{ \| {\cal S} x - \cA  (y) \|^{2}}\right)^{1/2},
\label{ravg}
\end{equation}
where $\Eval{\cdot}$ denotes the expected value taken with respect to the conditional measure $\muX$ introduced in Remark \ref{rem-cond-meas} (which is also Gaussian, due to well-known properties of normal measures). It follows that the optimal average estimate is immediately given by
\[
z\ped{o}\apex{av}=\cS x\apex{ls},
\]
for any $y\in Y$.
Moreover, in \cite[Chapter 6]{TrWaWo:88} it is proven that the optimal average radius does not depend on the measurement $y$, and it can be computed in closed form as
\[
r\apex{av}\ped{o}=r\apex{av}\ped{o}(y) = \sqrt{\mathrm{Trace}\left(\cS \left(\cI\tran \Sigma^{-1} \cI\right)^{-1}\cS\tran\right)}.
\]

For what concerns the probabilistic optimal estimate, we first remark that in the case of normally distributed noise, the definition of the probabilistic radius (\ref{rpr}) still applies, observing that the consistency set $\Ix$ defined in (\ref{set}) in this case is given by
\[
\Ix  \doteq \left\{ x \in X \,:\,  y = \cI x + \q ,\;\q \in \Real{n} \right\},
\]
and is unbounded. Hence, the ``discarded'' set $\cX_{\epsilon}$ in (\ref{rpr}) can be also unbounded. Note that this is not an issue, since  $\muX$ is defined over all $\Real{n}$, so that the measure of unbounded sets is well defined.

Similarly to the worst-case and the average settings, the optimality properties of the least-square solution still hold for the probabilistic setting. Indeed, in \cite[Chapter 8]{TrWaWo:88} it is proven that the optimal probabilistic estimate (to level $\epsilon$) for normal distributions is given by 
\[
z\ped{o}\apex{pr}=\cS x\apex{ls},
\]
for any $y\in Y$.
Closed-form solutions for the computation of the probabilistic radius $r\apex{pr}\ped{o}(\epsilon)$ are not available, and in \cite[Chapter 8]{TrWaWo:88}  the following upper bound is given
\[
r\apex{pr}\ped{o}(\epsilon,y)\le \sqrt{2\ln\frac{5}{\epsilon}}\,r\apex{av}\ped{o}, \quad 
\text{ for all } y\in Y.
\]
However, it is also observed that this bound is essentially sharp when the noise variance is sufficiently small.


\section{Numerical example}
\label{sec-examples}

\begin{figure}[!htb]
\centerline{
\includegraphics[width=8.5cm]{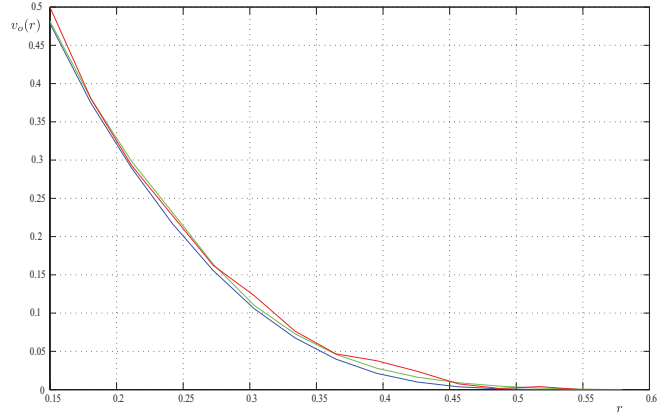}
}
\caption{Function $v\ped{o}(r)$ for the numerical example.
The blue line is the solution obtained computing the volume using the deterministic techniques discussed in Remark \ref{rem-volume}, the red line corresponds to the SPSA-algorithm and the green one is the SDP relaxation.
\label{fig-es-zoom}}
\end{figure}

As a numerical example, we consider a randomly generated instance of (\ref{meas})
with uniform distributed noise. In particular, $m=150$ random measurements of an unknown
$n=5$ dimensional vector were drawn taking 
\bea
\cI&=&\mathtt{round(20*rand(m,n)-10)} \label{ex-instance}\\
\q&=&\rho\mathtt{(2*rand(m,1)-1)}, \nonumber
\eea
with $\rho=5$, and considering as ``true'' parameters $x\apex{true}$ the unit vector.
The solution operator was chosen as
\[
\cS=
\left[
\begin{array}{ccccc}
    -5  &  10   &   -7& 0 & 0\\
     3  &   -4   &   7& 0 & 0\\
     2  &    6   &  4& 0 & 0
\end{array}
\right],
\]
leading to $z\apex{true}=\cS x\apex{true}= [-2\; 6\; 12]\tran$.
First, the optimal worst-case radius and the corresponding optimal solution have been computed by solving 10 linear programs
(corresponding to finding the tightest box containing the polytope $\cS\Ix$, see \cite{MilTem:85}). The computed worst-case optimal estimate is $z\ped{o}\apex{wc}=[
  -1.831\;
   5.839\;
   11.883
]\tran$ and the worst-case radius is
$r\ped{o}\apex{wc}(y)=0.5791$. 
\begin{figure}[!htb]
\centerline{
\includegraphics[width=11.5cm]{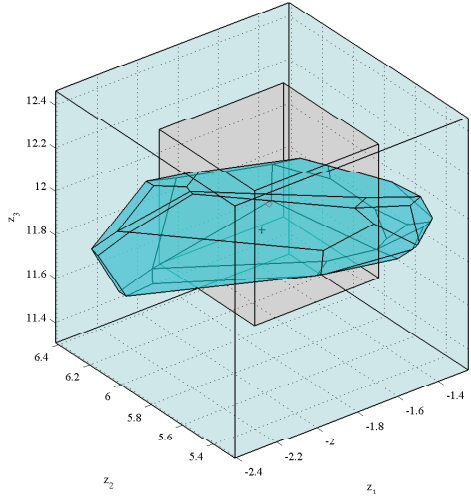}
}
\caption{Consistency set, and optimal ``box'' discarding a set of  measure $\epsilon=0.1$. The probabilistic optimal radius $r\ped{o}\apex{pr}(y,0.1)$ corresponds to the radius of this box. The center, denoted by a star, represents the optimal probabilistic  estimate $z\ped{o}\apex{pr}(y)$. \label{fig-es-pr}}
\end{figure}
Subsequently, in order to apply the proposed probabilistic framework, we fixed the accuracy level to $\epsilon=0.1$, and computed the probabilistic optimal radius and the corresponding optimal estimate according to definitions (\ref{rpr_opt}) and (\ref{prob-opt-est}).
In this case, we were still able to use the techniques discussed in Remark \ref{rem-volume} for computing \Pmax exactly. By employing a simple bisection search algorithm over $v\ped{o}(r)$, the probabilistic radius of information  was computed as $r\ped{o}\apex{pr}(y,0.1)=0.3074$.
The corresponding optimal probabilistic estimate is given by 
and $z\ped{o}\apex{pr}(0.1)=[
   -1.773\;
   5.869\;
  11.969
]\tran$. Note that the reduction in terms of radius of information is quite significant, being of the order of $50\%$. The meaning of our approach is well explained in Figure \ref{fig-es-pr}. Indeed, in this figure we see that we look for the optimal ``box'' discarding a set of probability measure $\epsilon=0.1$.
Note that, in this figure, the volume of the ``discarded set'' is clearly more than $10\%$ of the total volume. The reason of this is that  the probability of the discarded set is measured in the (five dimensional) space $X$.
Figure \ref{fig-es-zoom} shows a plot of the violation function $v_o(r)$
computed using the different techniques discussed in this paper. It can be observed that all methods provide very consistent results.

\begin{figure}[ht]
\centerline{
\includegraphics[width=22cm]{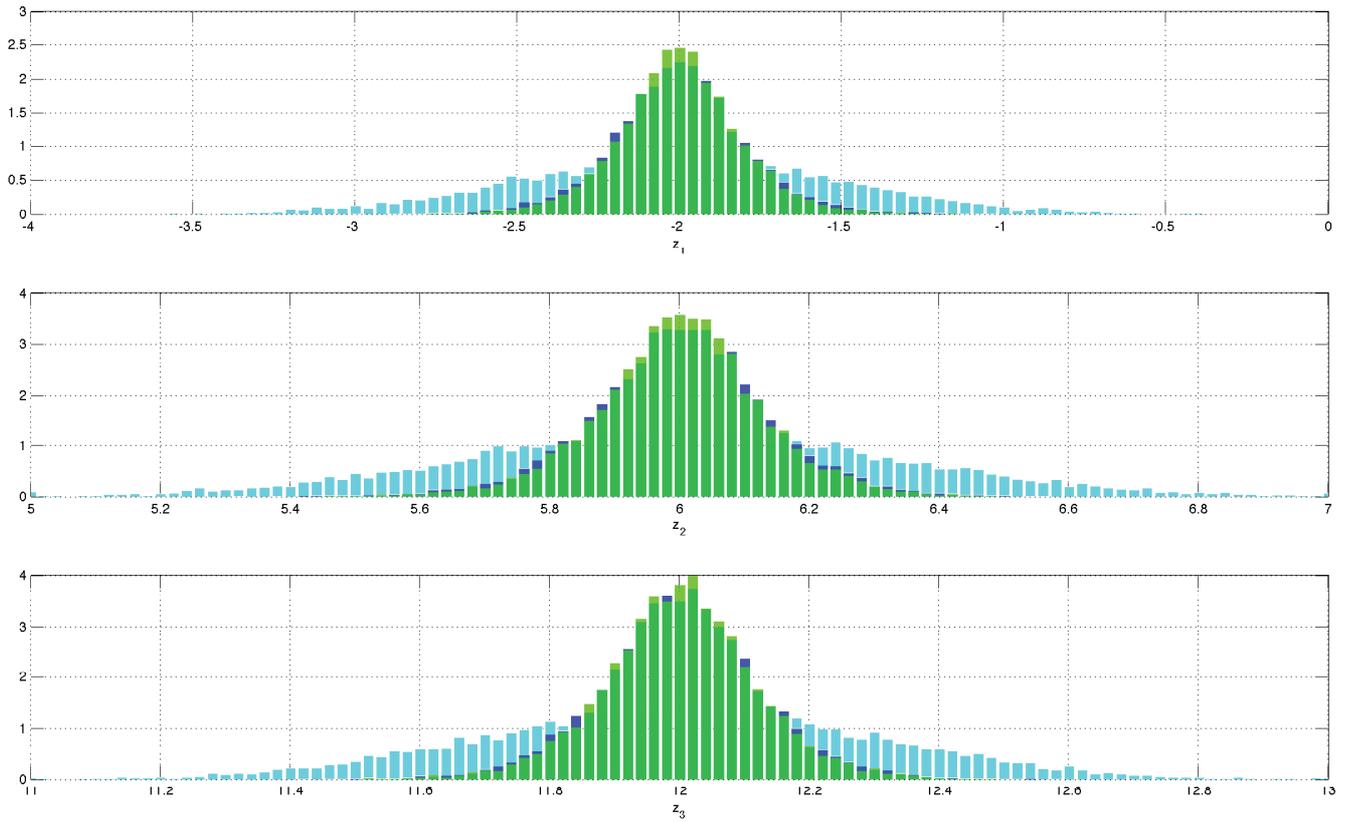}
}
\caption{Relative frequencies for least-square estimates (cyan), worst-case optimal estimate (blue) and probabilistic-optimal (green). To allow comparison, the histograms have been normalized so that so that the area under the bars is equal to one.
\label{fig-es-compare}}
\end{figure}

Then, to compare, we run $N=10,000$ random experiments, generating each time a difference instance of ({ex-instance}), and for each we computed the least-square estimate $z\apex{ls}=\cS x\apex{ls}$, the 
worst-case optimal estimate $z\ped{o}\apex{wc}$, and the probabilistic optimal estimate $z\ped{o}\apex{pr}$. Figure \ref{fig-es-compare} shows the (normalized) relative frequency histograms for the three estimates, while Table \ref{tabex} reportes the mean and variances
of the different estimates and their corresponding errors. It can be seen that both mean and variance of the error of the proposed probabilistic estimate are much smaller than the least-squares one, and also smaller than the worst-case one.

\begin{table}[htdp]
\label{tabex}
\caption{Mean and variance of least-squares, worst-case and probabilistic optimal estimates and corresponding errors}
\begin{center}
\scalebox{.95}{
\begin{tabular}{|c|c|c|}
\hline
\hline
& Mean & Variance\\
\hline
\hline
$z\apex{ls}$   & 
   -2.0035 \,   6.0017 \,  11.9993
& 
    0.2618   \, 0.1117 \,   0.0933
\\ \hline
$z\ped{o}\apex{wc}$ &  
   -2.0007  \,  6.0005 \,  12.0006
& 
    0.0420  \,  0.0183  \,  0.0151
\\ \hline
$z\ped{o}\apex{pr}$  &  
   -2.0008  \,  6.0004 \,  12.0004
& 
    0.0364  \,  0.0162  \,  0.0140
\\ 
\hline
$\|z\ped{o}\apex{ls}-z\apex{true}\|_{\infty}$   & 
   0.4764&    0.0762\\
\hline
$\|z\ped{o}\apex{wc}-z\apex{true}\|_{\infty}$   & 
    0.1864&    0.0148\\
\hline
$\|z\ped{o}\apex{pr} -z\apex{true}\|_{\infty} $ & 
    0.1725&    0.0134\\
\hline
\hline
\end{tabular}}
\end{center}
\label{default}
\end{table}%

\section{Conclusions}
\label{sec-conclusions}
This paper deals with the rapprochement between the stochastic and worst-case settings for system identification. The problem is formulated within the probabilistic setting of information-based complexity, and it is focused on the idea of discarding sets of small measure from the set of deterministic estimates. The paper establishes rigorous optimality properties of a trade-off curve, called \textit{violation function}, which  shows how the radius of information decreases as a function of the accuracy.  Subsequently, randomized and deterministic algorithms for computing the optimal violation function  have been presented. Their performance has been successfully tested on a numerical example.

\section{Acknowledgements}
The authors would like to thank Prof.\ Takayuki Wada for the enlightening discussions about stochastic optimization, and for clarifying the quasi-concavity properties in Theorem \ref{th-difference}, and Prof.\ Libor Vesely for clarifying properties of the Chebychev center of convex sets. The authors are also indebted to the anonymous reviewers for the numerous comments that allowed to improve the content of the paper.

\appendices

\section{Proof of Theorem \ref{theorem1}}
\label{app-theorem1}

Consider the transformation matrix  $T=\left[T_{1}\; T_{2}\right]$, where $T_{1}$  is an orthonormal basis of the column space of $\cI$
and  $T_{2}$ is an orthonormal basis of the null space of  $\cI\tran$.
Furthermore, define the linear transformation $\bar \q\doteq T\tran \q$, and
 the set $\bar \Q \doteq \left\{\bar \q \in\Real{m} \,:\, T^{-\!\top}\bar \q \in \Q\right\}$.
Then, if the random variable $\q$  is uniform on $\Q$,
the linearly transformed random variable $\bar \q$ is uniform on $\bar \Q$
(see e.g.\ \cite{Halmos:50}). 
Next, by multiplying equation (\ref{meas}) from the left by $T\tran\!\!$, and defining
$\bar \cI_{1}\doteq T_{1}\tran \cI$ and $\bar y_{i}\doteq T_{i}\tran y$,
$\bar \q_{i}\doteq T_{i}\tran \q$, $i=1,2$, we get
\beq
\label{qqq}
\left\{
\ba{lcll}
\bar \cI_{1}x &+ &\bar \q_{1} &= \bar y_{1}\\
&&\bar \q_{2} &= \bar y_{2},
\ea
\right.
\eeq
since, by construction,  $T_{2}\tran \cI =0$.
It follows from definition (\ref{set}) that a point $x\in X$ belongs to $\Ix$ if and only if there exists  
$\bar \q=\left[
\begin{smallmatrix}
\bar \q_{1}\\
\bar \q_{2}
\end{smallmatrix}
\right]\in \bar \Q$ such that  (\ref{qqq}) holds, i.e.\ if there exist $\bar \q_{1}$ in the  set
$
\bar \Q_{1}
\doteq
\left\{
\bar \q_{1} \in \Real{n}\,:\,
\left[
\begin{smallmatrix}
\bar \q_{1}\\
\bar y_{2}
\end{smallmatrix}
\right]\in\bar \Q
\right\}
$
for which $\bar \cI_{1}x + \bar \q_{1} = \bar y_{1}$.
Note that the set $\bar \Q_{1}$ represents the intersection of the set $\bar \Q$ with the hyperplane $
\left\{
\bar \q = \left[
\begin{smallmatrix}
\bar \q_{1}\\
\bar \q_{2}
\end{smallmatrix}
\right]\in\Real{m}
\,:\,\bar \q_{2}=\bar y_{2}\right\}$.
Since $\bar \q$ is uniform on $\bar \Q$, it is also uniform on any subset of $\bar \Q$, and in particular on this intersection set.
Hence, $\bar \q_{1}$ is uniformly distributed on $\bar \Q_{1}$.

Statement (i) is proved noting that, from (\ref{qqq}), an element $x\in\Ix\subset\Real{n}$ can be written as the  mapping of  $\bar \q_{1}\in\bar \Q_{1}\subset\Real{m}$
through the one-to-one affine transformation $x=\cI_{1}^{-1}(\bar \q_{1}-\bar y_{1})$. Since bijective linear transformations preserve uniformity \cite{PapPil:02}, it follows that the random variable $x$ is uniformly distributed on $\Ix$.

Point (ii) follows immediately from the fact that the image of a uniform density through a linear operator $\Real{n}\to\Real{s}$ with $s\le n$ is log-concave (see e.g.\ \cite{PapPil:02}).
\qed

\section{Proof of Theorem \ref{th-difference}}
\label{app-th-difference}

To prove point (i), we first 
%
%
consider equation  (\ref{vr-opt}). Recalling that $\muQ$ is the uniform measure over~$\Q$, we  write
\beas
\lefteqn{
v\ped{o}(r)
=\inf_{\cA}\frac{\vol{ \left\{x \in \Ix\,:\, \|\cS  x - \cA(y) \| > r \right\}}}{\vol{\Ix}}   }  \\
&=&\frac{1}{\vol{\Ix}}\inf_{z\ped{c}} \vol{
\left\{x \in \Ix \,:\, \|  \cS x 
- z\ped{c} \| > r 
\right\} } \nonumber \\
&=&\frac{1}{\vol{\Ix}}\inf_{z\ped{c}} \vol{ 
\left\{x \in \Ix \,:\,   x\not\in  \Cyl(z\ped{c},r)
\right\} } \nonumber \\
&=&\frac{1}{\vol{\Ix}}\inf_{z\ped{c}} \vol{ 
\Ix 
\setminus \Cyl(z\ped{c},r)}.
\label{Dx}
\eeas
Next, we note that this equation  can be rewritten as 
the following maximization problem
\beas
\label{vmaxD}
v\ped{o}(r)&=&1-\frac{1}{\vol{\Ix}}\, \sup_{z\ped{c}} \vol{\Ix\,  \cap\, \Cyl(z\ped{c},r)}\\
&=&1-\frac{1}{\vol{\Ix}}\, \sup_{z\ped{c}} \phi(z\ped{c},r). 
\eeas
The statement in (i) follows immediately considering that this optimization problem can be restricted to the set $\cH(r)$
where the intersection  is non-empty.
The existence of a global maximum is guaranteed because $\cH(r)$ is compact and the function $\phi(z\ped{c},r)$ is continuous in $z\ped{c}$.

\vskip .1in

To prove point (ii), we first show that $\cH(r)$ is convex. Begin by noting that 
\beq \begin{array}{rcl}
 z\ped{c} \in \cH(r) & \iff &    \cS\Ix \cap \mathcal{B}(z\ped{c},r) \not = \emptyset \\
 & \iff &  d(z\ped{c},S\Ix) \leq r,
 \end{array}
 \eeq
where the distance 
$d(z,A)$ of a point $z \in \Real{s}$ to a given set $A \subset  \Real{s} $ is defined as $d(z,A) \doteq \inf_{x \in A} \|z-x\|_{p}$.
Since  the distance function to convex sets is convex, see e.g.\ \cite[Section 3.2.5]{BoyVan:04},  it follows, from convexity of $\cS\Ix$, that $d(z,\cS\Ix)$ is also convex. Hence, given $z\ped{c}^1,z\ped{c}^2 \in \cH(r)$, it follows that
$d(\alpha z\ped{c}^1 + (1-\alpha)z\ped{c}^2,S\Ix) \leq  \alpha d(z^1\ped{c},S\Ix) + (1-\alpha)d(z\ped{c}^2,S\Ix) \leq r$
$ \Rightarrow$ $ \alpha z\ped{c}^1 + (1-\alpha)z\ped{c}^2 \in \cH(r)$.
Then, note that problem (\ref{Vstar}) corresponds to maximizing the volume of the intersection $\Phi(z\ped{c},r)$ between the two convex sets $\Ix$ and $\Cyl(z\ped{c},r)$. One of them, $\Ix$,  is fixed, while the other one is the set obtained translating the cylinder 
$\Cyl(0,r)$ by $\left[\ba{c} \tS^{-1}z\ped{c}\\ 0 \ea\right]$. Similar problems have been studied in convex analysis, see for instance \cite{Zalgaller:01}.
In particular, the proof of  continuity follows closely the proof of Lemma 4.1 in \cite{FukUno:07}.
That is, consider an arbitrary direction $\xi\in\Real{s}$, and let $V_{\cC}$ be the volume of the set obtained projecting 
$\Cyl(0,r)$ to the hyperplane normal to $\xi$. Then, for any $\epsilon>0$, we have that the difference between the volume of $\Phi(z\ped{c},r)$ and $\phi(z\ped{c}+\epsilon\xi,r)$ is bounded by $\epsilon V_{\cC}\|\xi\|$. Hence, $\phi(z\ped{c},r)-\phi(z\ped{c}+\epsilon\xi,r)$
converges to zero for $\epsilon\to 0$, thus proving continuity.\\
To prove quasi-concavity, consider two points $z_{1},z_{2}\in\cH(r)$ such that $\phi(z_{1},r)>\phi(z_{2},r)$. Consider then a point $z_{\alpha}\doteq \alpha z_{1}+ (1-\alpha) z_{2}$ where
$\alpha\in(0,\,1)$. From convexity of $\cH(r)$ it follows that $z_{\alpha}\in\cH(r)$. 
Then, the following chain of inequalities holds
\bea
\lefteqn{\phi(z_{\alpha},r)^{1/n} =\phi(\alpha z_{1}+ (1-\alpha) z_{2})^{1/n}} \label{ssqc1}\\
&\ge&   \vol{\alpha \Phi(z_{1},r) + (1-\alpha) \Phi(z_{1},r)}^{1/n}\label{ssqc2}\\
&\ge&   \alpha\vol{\Phi(z_{1},r)}^{1/n} + (1-\alpha)\vol{\Phi(z_{1},r)}^{1/n}\label{ssqc3}\\
&=&   \alpha\phi(z_{1},r)^{1/n} + (1-\alpha)\phi(z_{1},r)^{1/n}\label{ssqc4}\\
&>& \alpha\phi(z_{2},r)^{1/n} + (1-\alpha)\phi(z_{1},r)^{1/n}\label{ssqc5}\\
&=& \phi(z_{2},r)^{1/n}
\eea
where (\ref{ssqc2}) follows  from \cite[Theorem 1]{Zalgaller:01},
(\ref{ssqc3}) follows  from the Brunn-Minkowski inequality for convex analysis
\cite{Schneider:93} and 
(\ref{ssqc3}) follows  from the hypothesis that $\phi(z_{1},r)>\phi(z_{2},r)$.
From this chain of inequalities, we have $\phi(z_{\alpha},r)>\phi(z_{2},r)$, which implies semi-strict quasi-concavity\footnote{A  function $f$ defined on a convex set
$A\in\Real{n}$ is semi-strictly quasi-concave if
$f(y)<f(\alpha x + (1-\alpha)y)$
holds for any  $x,y\in A$ such that $f(x)>f(y)$ and $\alpha\in (0,\,1)$.}.
Combining continuity and semi-strict quasi-concavity one finally gets quasi-concavity \cite{DiAvZa:81}.
\vskip .1in

To prove point (iii), we note that $v\ped{o}(r)$ is right continuous and non-increasing if and only if
$\phi\ped{o}(r)$ is upper semi-continuous and non-decreasing. 
To show upper semi-continuity of the supremum value function $\phi\ped{o}(r)$, consider the radius $\bar r= r\apex{wc}\ped{o}(y)$, which is nonzero
since $\cH(\bar r)$ is assumed non-empty. 
Then, from point (ii) it follows that, for any $\bar z\in \cH(\bar r)$, the upper level set 
$F(\bar z)\doteq\left\{z\in\cH(\bar r)\,:\, \phi(z,\bar r) \ge\phi(\bar z,\bar r) \right\}$
is strictly convex. Hence, the function $\phi(\cdot,r)$ is quasi-convex, continuous and satisfies the boundedness condition defined in \cite{Klatte:97}.
Then, upper semi-continuity of  $\phi\ped{o}(r)$ follows from direct application of  \cite[Theorem 2.1]{Klatte:97}. 
Finally, to show that $\phi\ped{o}(r)$ is non-decreasing, take $0<r_{1}<r_{2}$ and denote $z_{1}$ and $z_{2}$ be the optimal solutions corresponding to
$\phi\ped{o}(r_{1})$ and $\phi\ped{o}(r_{2})$, respectively. It follows that 
\beq
\label{frg1}
\phi(z_{\mathrm{o}1},r_{2})\le\phi(z_{\mathrm{o}2},r_{2})=\phi\ped{o}(r_{2}), 
\eeq
since $z_{\mathrm{o}2}$ is the point where the maximum is attained. On the other hand, from definition (\ref{Phi}) and $r_{1}<r_{2}$ we have 
\[
\Phi(z_{\mathrm{o}1},r_{1})= \left(\Ix\,  \cap\, \Cyl(z_{\mathrm{o}1},r_{1})\right) \,
\subseteq\, \left( \Ix\,  \cap\, \Cyl(z_{\mathrm{o}1},r_{2})\right)
\]
and hence 
\beq
\label{frg2}
\phi\ped{o}(r_{1})=\phi(z_{\mathrm{o}1},r_{1}) \le \phi(z_{\mathrm{o}1},r_{2}).
\eeq
Combining (\ref{frg1}) and (\ref{frg2}) it follows $\phi\ped{o}(r_{1})\le\phi\ped{o}(r_{2})$.
\qed
\bibliographystyle{plain}
\bibliography{/Users/fabriziodabbene/Dropbox/Biblio/Frugi-biblio}

\end{document}